\documentclass[graybox]{svmult}

\usepackage{mathptmx}       % selects Times Roman as basic font
\usepackage{helvet}         % selects Helvetica as sans-serif font
\usepackage{courier}        % selects Courier as typewriter font
\usepackage{type1cm}        % activate if the above 3 fonts are
\usepackage{makeidx}         % allows index generation
\usepackage{graphicx}        % standard LaTeX graphics tool
\usepackage{multicol}        % used for the two-column index
\usepackage[bottom]{footmisc}% places footnotes at page bottom

\usepackage{amssymb}
\makeindex             % used for the subject index
\begin{document}

\title*{Snapshots of Conformal Field Theory}
\author{Katrin Wendland}
\institute{Katrin Wendland \at Mathematics Institute, Freiburg University, Eckerstr. 1, 79104 Freiburg i. Br., Germany,\\ \email{katrin.wendland@math.uni-freiburg.de}}
\maketitle
\abstract{ In snapshots, this exposition introduces conformal field theory, with a 
focus on those perspectives that are relevant for interpreting superconformal field theory by 
Calabi-Yau geometry. It includes a detailed discussion of the elliptic genus as an invariant which certain 
superconformal field theories share with the Calabi-Yau manifolds. K3 theories are (re)viewed as 
prime examples of superconformal field theories where geometric interpretations are known. A final 
snapshot addresses the K3-related Mathieu Moonshine phenomena, where a lead role is predicted for the chiral de 
Rham complex. }
%
%the abstract should be 10--15 lines long
%
%
\section{Introduction}\label{intro}
Conformal quantum field theory (CFT) became popular in physics thanks to the work by Belavin, Polyakov
and Zamolodchikov. 
In their seminal paper \cite{bpz84}, on the one hand, they lay the mathematical foundations of axiomatic CFT, 
and on the other hand, they show the physical significance of CFT for surface phenomena in statistical physics by 
describing certain phase transitions of second order through CFT.

Another common source of conformal field theories is string theory, which is many theoreticians' favorite candidate for 
the unification of all interactions, including gravity. Here, particles are described by strings that move in some
potentially complicated background geometry. The string dynamics are governed by a so-called non-linear sigma model, 
such that conformal invariance yields the string equations of motion. The quantum field theory living on the worldsheet 
of the string then is a CFT. This implies deep relations between CFT and geometry, which have already led to a number of 
intriguing insights in geometry, demanding for a more resilient bridge between mathematics and physics.

For example, in the early 90s mirror symmetry provided a first success story for the interaction between 
mathematics and physics in the context of CFT \cite{lvw89,grpl90,cogp91,cls90}.
However, a rigorous approach to those types of CFTs that are relevant for such deep insights in algebraic 
geometry was not available, at the time. As a result, the interaction between mathematics and physics in many 
cases amounted to a rather imbalanced division of work, where theoretical physicists provided the most amazing 
predictions and left them to the mathematicians for a proof, who in turn successfully detached their theories from 
their origins in physics.

With the advent of Monstrous Moonshine \cite{th79a,th79b,cono79,flm84,bo86,ga06},
and with Borcherds' Fields Medal in 1998 ``for his contributions to algebra, the theory of automorphic forms, 
and mathematical physics, including the introduction of vertex algebras and Borcherds' Lie algebras, the proof of the 
Conway-Norton moonshine conjecture and the discovery of a new class of automorphic infinite products''
\cite{fields98},  the subject of conformal field theory, per se, began to become more popular in mathematics. 
Indeed, the comparatively new notion of vertex algebras provided a rigorous mathematical foundation to the 
most basic ingredients of conformal quantum field theory and thereby offered a viable approach to CFT for mathematicians.
Nevertheless, the quest to fill the gap between abstract mathematical approaches to CFT and those types of models that are of
interest in physics, and that are relevant for deeper insights in algebraic and enumerative geometry, 
has not yet been completed. The present work attempts to make a contribution to this quest.\\

Since this exposition can certainly only provide some snapshots of CFT, it has to follow a subjective 
selection and presentation of material. The guiding principle is the conviction that on the one hand, 
the foundation of the discussion has to be a mathematically rigorous definition of CFT, which is independent 
of string theory, while on the other hand, those predictions from CFT which affect the geometry of Calabi-Yau 
manifolds are among the most intriguing ones. To state and understand the latter, one needs to work with a 
mathematical formulation of CFT which allows to make contact with the non-linear sigma models in physics, 
thus sadly excluding a number of popular approaches to CFT. Moreover, the discussion is restricted to so-called 
two-dimensional Euclidean unitary CFTs.
\\

\noindent
In more detail, this work is structured as follows:

Section \ref{cft} provides a definition of some of the ingredients of CFT. The conformal
vertex algebras serve as our point of entry in Section \ref{vertexalgebras}, since this part of
CFT is probably the most natural for mathematicians. We proceed in Section \ref{cftdef} 
by listing the crucial ingredients that underlie a definition of superconformal field theory, 
along with additional required properties. The presentation makes no claim for completeness, but 
according to our declared conviction, we focus on those aspects that are relevant for the discussion of geometric 
interpretations as introduced later. This in particular restricts our attention to the so-called $N=(2,2)$ 
superconformal field theories with space-time supersymmetry. A useful class of examples, which is well 
understood, is given by the toroidal $N=(2,2)$ superconformal field theories presented in 
Section \ref{toroidalcfts}. We summarize the definition and properties of the chiral de Rham complex in 
Section \ref{chiderham}, as an example of a sheaf of conformal vertex algebras on an arbitrary 
smooth algebraic variety, which thus provides a link between standard ingredients of CFT and 
geometric quantities. Since this link is not entirely understood to the very day, for clarification, 
our discussion rests on the special role of the elliptic genus as an invariant that certain 
superconformal field theories share with the Calabi-Yau manifolds\footnote{Disclaimer: in this
work, all Calabi-Yau manifolds are compact, by definition.}, 
as is discussed in some detail in Section \ref{ellgen}.

The elliptic genus is also crucial for our definition of K3 theories in Section \ref{k3cft}. 
This class of CFTs deserves some attention, as it provides the only examples of non-linear sigma 
models on Calabi-Yau manifolds other than tori, where at least there are precise predictions on the global form 
of the moduli space, implying some very explicit relations between quantities in geometry and CFT. 
We motivate the definition of K3 theories in detail, and we summarize some of the known 
properties of these theories. In particular, Proposition \ref{twoellgens} recalls the dichotomy of $N=(2,2)$ 
superconformal field theories at central charges $c=6,\,\overline c=6$ with space-time supersymmetry and integral 
$U(1)$-charges. Indeed, these theories fall into two classes, namely the toroidal 
and the K3 theories. Thus Proposition \ref{twoellgens} is the conformal field theoretic counterpart of the 
classification of 
 Calabi-Yau $2$-manifolds into complex two-tori, on the one hand, 
and K3 surfaces, on the other. Our proof \cite[\S7.1]{diss}, which is little known, is summarized in the Appendix.

The final Section \ref{ellk3} is devoted to recent developments in the study of K3 theories, 
related to the mysterious phenomena known as Mathieu Moonshine. We recall the route to 
discovery of these phenomena, which also proceeds via the elliptic genus. We  offer some 
ideas towards a geometric interpretation, arguing that one should expect the chiral de 
Rham complex to be crucial in unraveling the Mathieu Moonshine mysteries. The section closes with an open conjecture, 
which is related to Mathieu Moonshine, which however is formulated  neither alluding to moonshine nor to CFT, 
and which therefore is hoped to be of independent interest.
\section{Ingredients of conformal field theory}\label{cft}
The present section collects ingredients of conformal field theory (CFT),
more precisely of \textsc{two-dimensional Euclidean unitary conformal 
field theory}. These adjectives translate into the properties of the underlying quantum field theory as follows:
first, all fields  are parametrized on a
\textsc{two-dimensional worldsheet}, which comes equipped with a \textsc{Euclidean metric}.
Second, the fields transform {covariantly} under \textsc{conformal
maps} between such worldsheets. Furthermore, the space of states in such a CFT is 
equipped with a {positive
definite metric}, with respect to which the infinitesimal conformal transformations act \textsc{unitarily}.

We begin by describing
the simplest fields in our CFTs  in terms of the so-called \textsc{vertex algebras}
in Section \ref{vertexalgebras}. Next,
Section \ref{cftdef} summarizes a definition of conformal field theory, 
with the  toroidal conformal field theories presented as a class of examples
in Section \ref{toroidalcfts}.
In Sections
\ref{ellgen} and \ref{chiderham} the related notions of the elliptic genus and the chiral de Rham complex are discussed
in the context of superconformal field theories. 
As such, the present section collects ingredients of CFT, with a focus on
some of those ingredients that are
under investigation to the very day.
\subsection{Conformal and superconformal vertex algebras}\label{vertexalgebras}
We begin by recalling the 
notion of \textsc{fields}, following \cite{ka96}.
The theory is built on the earlier results \cite{lewi78,frka80,bo86},  
see also \cite{frbe04} for a very readable exposition. This definition
 is most convenient,
because it naturally implements the representation theory inherent to CFTs.
As we shall see at the end
of this section, for
the chiral states of CFTs it also allows a straightforward definition of the $n$-point functions.
\begin{svgraybox}\vspace*{-1.5em}
\begin{definition}\label{fielddef}
\hspace*{\fill}

\noindent
Consider a $\mathbb C$-vectorspace $\mathbb H$.
\begin{itemize}
\item
 $\mathbb H[\hspace{-.1em}[ z_1^{\pm1},\ldots,z_n^{\pm1}]\hspace{-.1em}]$
denotes the vectorspace of formal power series
$$
v(z_1,\ldots,z_n) 
= \sum_{i_1,\ldots,i_n\in{\mathbb Z}} \widehat v_{i_1,\ldots,i_n} z_1^{i_1}\cdots z_n^{i_n},
\quad \widehat v_{i_1,\ldots,i_n}\in \mathbb H.
$$
\item
For 
$A\in\mbox{End}_{\mathbb C}({\mathbb H})[\hspace{-.1em}[ z_1^{\pm1},\ldots,z_n^{\pm1}]\hspace{-.1em}]$,
and for
$\alpha\in{\mathbb H}^\ast:=\mbox{Hom}_{\mathbb C}({\mathbb H},{\mathbb C})$ and $v\in\mathbb H$, 
we set
$$
\langle\alpha,A(z_1,\ldots,z_n) v\rangle
:= 
\sum_{i_1,\ldots,i_n\in{\mathbb Z}} 
\langle\alpha, \widehat A_{i_1,\ldots,i_n} v\rangle z_1^{i_1}\cdots z_n^{i_n}
\quad\in \mathbb C[\hspace{-.1em}[ z_1^{\pm1},\ldots,z_n^{\pm1}]\hspace{-.1em}],
$$
where on the right hand side,
$\langle\cdot,\cdot\rangle$ denotes the natural pairing between $\mathbb H^\ast$ and $\mathbb H$.
\item 
If $A(z)\in \mbox{End}_{\mathbb C}({\mathbb H})[\hspace{-.1em}[ z^{\pm1}]\hspace{-.1em}]$
with $A(z)=\sum_n \widehat A_n z^n$, then $\partial A$ denotes
the formal derivative of $A$,
$$
\partial A(z) =\sum_{n\in\mathbb Z} n\widehat A_n z^{n-1}
\quad \in \mbox{End}_{\mathbb C}({\mathbb H})[\hspace{-.1em}[ z^{\pm1}]\hspace{-.1em}].
$$
\item
A formal power series $A(z)\in \mbox{End}_{\mathbb C}({\mathbb H})[\hspace{-.1em}[ z^{\pm1}]\hspace{-.1em}]$ is called
a \textsc{field} on $\mathbb H$ if $A(z)=\sum_n \widehat A_n z^n$ obeys
$$
\forall v\in{\mathbb H}\colon\quad \exists N\in{\mathbb Z} \mbox{ such that }
\widehat A_nv =0\;\;\forall n< N.
$$
The endomorphisms $\widehat A_n$ are called the \textsc{modes}
of the field $A$. 
\end{itemize}
\end{definition}
\vspace*{-1em}
\end{svgraybox}
In other words, if $A$ is a field on ${\mathbb H}$, then for every 
$v\in{\mathbb H}$ the expression $A(z)v = \sum_n (\widehat A_n v) z^n$
is a formal Laurent series with coefficients in ${\mathbb H}$ and with only finitely many
non-zero contributions $(\widehat A_nv)z^n$ with $n<0$. In the context
of CFTs
one can introduce a completion $\overline{\mathbb H}$ of ${\mathbb H}$ with respect
to an appropriate topology and then for every $z\in{\mathbb C}^\ast$ 
view $A(z)$ as a linear
operator from ${\mathbb H}$ to $\overline{\mathbb H}$, see for example
\cite[\S1.2.1]{frbe04}. 
Accordingly, we call a field $A(z)=\sum_n\widehat A_n z^n$ \textsc{constant} if $\widehat A_n=0$
for all $n\neq0$. Similarly, if for $v\in\mathbb H$ we have $\widehat A_nv=0$
for all $n<0$, then we say that \textsc{$A(z)v$ is well-defined in $z=0$}, and
$A(z)v_{\mid z=0}=\widehat A_0v$.
Note that 
a field $A$ according to Definition \ref{fielddef} can 
be viewed as an operator valued distribution, as usual in quantum field theory. Indeed, 
by means of the residue,
$A(z)$ yields a linear map from complex polynomials into $\mathbb H$. 
By definition, the space $\mathbb H$ carries a representation of the Lie algebra 
generated by the modes of every field on $\mathbb H$, with the Lie bracket that is
inherited from $\mbox{End}_{\mathbb C}(\mathbb H)$, namely the commutator.
\\

Let us now consider two fields $A,\, B$ on $\mathbb H$. 
While the expressions $A(z)B(w)$ and $B(w)A(z)$ make sense as formal power series 
in $\mbox{End}_{\mathbb C}({\mathbb H})[\hspace{-.1em}[ z^{\pm1}, w^{\pm1}]\hspace{-.1em}]$,
a priori it is impossible to interpret them as fields. 
In general, we expect singular behavior for the coefficients 
when we insert $w= z$, and in fact the form of this singularity captures the most important 
aspects of CFT.
Here,
 the notions of \textsc{locality} and \textsc{normal ordered products} come to aid:
\begin{svgraybox}\vspace*{-1.5em}
\begin{definition}\label{locality}
\hspace*{\fill}
\begin{enumerate}
\item
Let $\partial_w$ denote the formal derivative with respect to $w$ in
$\mathbb C[\hspace{-.1em}[z^{\pm1}, w^{\pm1}]\hspace{-.1em}]$.
On $\mathbb C[\hspace{-.1em}[ z, w]\hspace{-.1em}][z^{-1}, w^{-1}, (z-w)^{-1}]$, we define
the $\mathbb C[\hspace{-.1em}[ z, w]\hspace{-.1em}][z^{-1}, w^{-1}]$-linear operators 
$\iota_{z>w}$ and $\iota_{w>z}$ 
into $\mathbb C[\hspace{-.1em}[z^{\pm1}, w^{\pm1}]\hspace{-.1em}]$
with
\begin{eqnarray*}
\mbox{ for } k\in\mathbb N\colon\quad
\iota_{z>w} \left( k!(z-w)^{-k-1} \right) &=& {\partial_w^k} {1\over z}\sum_{n=0}^\infty \left( {w\over z}\right)^n, \quad\\
\iota_{w>z} \left( k!(z-w)^{-k-1} \right) &=& - {\partial_w^k} {1\over w}\sum_{n=0}^\infty \left( {z\over w}\right)^n
.
\end{eqnarray*}
\item
Fields $A,\, B$ on $\mathbb H$
are called \textsc{local with respect to each other} if there exist a so-called
\textsc{normal ordered product} 
$:\!\!A(z)B(w)\!\!\!: \in \mbox{End}_{\mathbb C}({\mathbb H})[\hspace{-.1em}[ z^{\pm1}, w^{\pm1}]\hspace{-.1em}]$
and fields $X_0,\ldots, X_{N-1}$ and $:\!AB\!\!:$
on $\mathbb H$, such that for every $\alpha\in\mathbb H^\ast$ and $v\in\mathbb H$,
\begin{itemize}
\item
we have $\langle\alpha, :\!A(z)B(w)\!\!: v\rangle\in \mathbb C[\hspace{-.1em}[ z, w]\hspace{-.1em}][z^{-1}, w^{-1}]$,
\item
in $\mbox{End}_{\mathbb C}({\mathbb H})[\hspace{-.1em}[ z^{\pm1}]\hspace{-.1em}]$, we have
$:\!AB\!\!:\!\!(z)\;=\;:\!A(z)B(w)\!\!:_{\mid w=z}$,
\item
in $\mathbb C[\hspace{-.1em}[ z, w]\hspace{-.1em}][z^{-1}, w^{-1}, (z-w)^{-1}]$, we have
\begin{eqnarray*}
 \langle \alpha, A(z)B(w) v\rangle 
&=&  \iota_{z>w} \left(\sum_{j=0}^{N-1} {\langle\alpha, X_j(w) v\rangle \over (z-w)^{j+1}} \right)+ \langle\alpha, :\!A(z)B(w)\!\!: v\rangle ,\\
\langle \alpha, B(w)A(z) v\rangle&=& \iota_{w>z}   \left(\sum_{j=0}^{N-1} {\langle\alpha, X_j(w) v\rangle \over (z-w)^{j+1}}\right) + \langle\alpha, :\!A(z)B(w)\!\!: v\rangle .
\end{eqnarray*}
\end{itemize}
As a shorthand notation one writes the so-called \textsc{operator product expansion (OPE)}
$$
A(z)B(w) \sim \sum_{j=0}^{N-1} {X_j(w)\over (z-w)^{j+1}},
$$
where  contributions that are regular at $z=w$  may be omitted on the right hand side
at will.
\end{enumerate}
\end{definition}\vspace*{-1em}
\end{svgraybox}
For  the special fields that feature in CFTs, 
the formal power series in the above definition yield convergent functions in 
complex variables $z$ and $w$ on appropriate domains in $\mathbb C$. Then
 the operators $\iota_{z>w}$ and $\iota_{w>z}$ implement the
Taylor  expansions about $z=w$ in the domains $|z|>|w|$ and $|w|>|z|$, respectively. 
We therefore refer to these operators as \textsc{(formal) 
Taylor expansions}.
The OPE thus captures the singular behavior of the expressions
$\langle \alpha, A(z)B(w) v\rangle$ when $z\sim w$, where locality of the fields $A$ and $B$
with respect to each other restricts the
possible singularities to poles at $z=w$. For $[A(z),B(w)]:=A(z)B(w)-B(w)A(z)
\in\mbox{End}_{\mathbb C}({\mathbb H})[\hspace{-.1em}[ z^{\pm1}, w^{\pm1}]\hspace{-.1em}]$,
the observation that, in general,
$\langle \alpha, [A(z),B(w)] v\rangle$ does not vanish, accounts for the fact that $(z-w)^{-1}$ and
its derivatives have different Taylor expansions in the domains $|z|>|w|$ and $|w|>|z|$, respectively.  
Hence the modes of
the fields $X_j$ in the OPE encode the commutators $[\widehat A_n,\widehat B_m]$ of the modes of 
$A$ and $B$. 

The Definition \ref{locality} of the normal ordered product $:\!A(z)B(w)\!\!:$ of two fields $A,\, B$ on $\mathbb H$ 
yields $:\!A(z)B(w)\!\!:=A_+(z)B(w)+B(w)A_-(z)$ if $A(z)=A_+(z)+A_-(z)$, 
where $A_+(z):=\sum_{n\geq0}\widehat A_n z^n$ and $A_-(z):=\sum_{n<0}\widehat A_n z^n$. 
Hence our definition of normal ordered product
amounts to a choice in decomposing $A(z)=A_+(z)+A_-(z)$ as stated,
which accrues from the choice of decomposing the formal power series
$$
\sum_{m=-\infty}^\infty z^m w^{-m-1} = \iota_{z>w}\left( (z-w)^{-1}\right) - \iota_{w>z}\left( (z-w)^{-1}\right)
\quad\in \mathbb C[\hspace{-.1em}[ z^{\pm1}, w^{\pm1}]\hspace{-.1em}].
$$
In the context of superconformal field theories, these notions are generalized to
include \textsc{odd} fields; if both $A$ and $B$ are odd, then locality amounts to
\begin{eqnarray*}
 \langle \alpha, A(z)B(w) v\rangle 
&=& \iota_{z>w} \left( \sum_{j=0}^{N-1} {\langle\alpha, X_j(w) v\rangle \over (z-w)^{j+1}}\right) + \langle\alpha, :\!A(z)B(w)\!\!: v\rangle ,\\
-\langle \alpha, B(w)A(z) v\rangle&=& \iota_{w>z} \left(\sum_{j=0}^{N-1} {\langle\alpha, X_j(w) v\rangle \over (z-w)^{j+1}} \right)+ \langle\alpha, :\!A(z)B(w)\!\!: v\rangle  ,
\end{eqnarray*}
abbreviated by the same OPE as in Definition \ref{locality},
and the bracket $[\cdot,\cdot]$ in the above argument is replaced by 
a superbracket with $[A(z),B(w)]=A(z)B(w)+B(w)A(z)$
for odd fields $A,\,B$.
The space $\mathbb H$, accordingly, furnishes a representation of the super-Lie algebra
generated by the modes of the fields on $\mathbb H$.\\

\noindent
The following list of examples  provides some basic fields in the simplest CFTs:
\begin{example}\label{U1current}
[\textsl{${U(1)}$-current}]\hspace*{\fill}

\noindent
We consider the complex Lie algebra $\mathcal A$ with $\mathbb C$-vectorspace basis 
$\left\{C;\;a_n,\, n\in\mathbb Z\right\}$, where $C$ is a central element and the
Lie bracket obeys
$$
\forall m,\, n\in\mathbb Z\colon\quad
[a_n,a_m] = m\delta_{n+m,0}\cdot {C\over3} .
$$
Choose some  $c\in\mathbb R$ and
let $\mathbb H$ denote the $\mathcal A$-module which under the $\mathcal A$-action 
is generated by a single 
non-zero vector $\Omega$, with submodule of relations generated by
$$
a_n\Omega = 0\quad \forall n\leq 0,\qquad C\Omega = c\Omega.
$$ 
The space $\mathbb H$ can be viewed as polynomial ring in the $a_n$ with $n>0$.
One then checks that the so-called \textsc{$U(1)$-current}
$$
J(z):= \sum_{n=-\infty}^\infty a_n z^{n-1}
$$
is a well-defined field on $\mathbb H$ which obeys the OPE
$$
J(z) J(w) \sim {c/3\over (z-w)^2}.
$$
In particular, $J$ is local with respect to itself.
Here and in the following, a constant field which
acts by multiplication by $\Lambda\in\mathbb C$ on $\mathbb H$ is simply denoted by $\Lambda$.
\end{example}
\begin{example}\label{Virasoro}
[\textsl{Virasoro field}]\hspace*{\fill}

\noindent
For the $U(1)$-current $J$ on the vectorspace $\mathbb H$
introduced in the previous example, assume $c\neq0$ and
let $T(z):= {3\over2c} :\!JJ\!\!:\!\!(z)$. One checks that
with $c^\bullet=1$
this field on $\mathbb H$ obeys the OPE
\begin{equation}\label{Virasorofield}
T(z)T(w) \sim  {c^\bullet\slash2\over (z-w)^4} + {2T(w)\over(z-w)^2}+ {\partial T(w)\over z-w},
\end{equation}
which for the modes of $T(z)=\sum_n L_n z^{n-2}$ translates into
\begin{equation}\label{Virasoroalgebra}
\forall n,\,m\in\mathbb Z\colon\quad
[L_n,L_m] = (m-n) L_{m+n} + \delta_{n+m,0}\, 
{\textstyle{c^\bullet\over12}}\, m(m^2-1).
\end{equation}
The above equation (\ref{Virasoroalgebra}) defines the \textsc{Virasoro algebra} at central charge
$c^\bullet$, whose underlying vectorspace has $\mathbb C$-vectorspace basis 
$\left\{c^\bullet;\;L_n,\, n\in\mathbb Z\right\}$. 
This Lie algebra is the central extension by $\mbox{span}_{\mathbb C}\{c^\bullet\}$ of the
Lie algebra of infinitesimal conformal transformations of the punctured 
Euclidean plane $\mathbb C^\ast$.
\end{example}
\begin{example}\label{bcbetagamma} 
[\textsl{$bc-\beta\gamma$-system}]\hspace*{\fill}

\noindent
Let $D\in\mathbb N$, and consider the super-Lie algebra $\mathcal A_D$ with 
$\mathbb C$-vectorspace basis 
$\left\{C;\right.$ $\left.a_n^i,\,b_n^i,\,\varphi_n^i,\,\psi_n^i,\,
\, n\in\mathbb Z,\,i\in\{1,\ldots,D\}\right\}$, where
the $a_n^i,\,b_n^i$ and the central element $C$ are even, while the 
$\varphi_n^i,\,\psi_n^i$ are odd, and the only non-vanishing basic super-Lie brackets
are
\begin{equation}\label{bcbgalgebra}
\begin{array}{rcl}
\forall m,\, n\in\mathbb Z,\; i,\,j \in\{1,\ldots,D\}\colon\quad
[a_n^i,b_m^j] &=& \delta^{i,j}\delta_{n+m,0}\cdot {C},\quad\\[3pt]
\{\psi_n^i,\varphi_m^j\} &=& \delta^{i,j}\delta_{n+m,0}\cdot {C}.\quad
\end{array}
\end{equation}
Here, $\{\cdot,\cdot\}$ denotes the super-Lie bracket between odd elements of 
$\mathcal A_D$, as is customary in the physics literature. 
Let $\mathbb H$ denote the $\mathcal A_D$-module which under the $\mathcal A_D$-action
is generated by a single non-zero vector $\Omega$, with submodule of relations generated by
\begin{eqnarray*}
\forall n\leq 0,\, m<0,\; i,\,j \in\{1,\ldots,D\}\colon\quad
a_n^i\Omega \;=\; 0,\;\; 
b_m^j\Omega &=& 0,\; \\
\psi_n^i\Omega \;=\; 0,\;\; 
\varphi_m^j\Omega &=& 0;\; \;
\quad C\Omega = \Omega.
\end{eqnarray*}
Generalizing Examples \ref{U1current} and \ref{Virasoro} above, one checks 
that 
\begin{eqnarray*}
a^i(z)\;:=\; \sum_{n=-\infty}^\infty a_n^i z^{n-1},\quad
b^i(z)&:=& \sum_{m=-\infty}^\infty b_m^i z^{m},\quad\\
\psi^i(z)\;:=\; \sum_{n=-\infty}^\infty \psi_n^i z^{n-1},\quad
\varphi^i(z)&:=& \sum_{m=-\infty}^\infty \varphi_m^i z^{m},\quad
i\in\{1,\ldots,D\}
\end{eqnarray*}
defines pairwise local fields $a^i,\, b^i,\, \psi^i,\, \varphi^i$ on $\mathbb H$.
Moreover, one finds the OPEs
$$
a^i(z) b^j(w) \sim {\delta^{i,j}\over z-w},\quad
\varphi^i(z) \psi^j(w) \sim {\delta^{i,j}\over z-w}\quad
\forall\;i,\,j \in\{1,\ldots,D\},
$$
while all other basic OPEs vanish,
and the field
\begin{equation}\label{topT}
T^{\rm top}(z):= \sum_{j=1}^D \left(  :\!\partial b^ja^j\!\!:\!\!(z)+ :\!\partial\varphi^j\psi^j\!\!:\!\!(z)\right)
\end{equation}
is a Virasoro field obeying (\ref{Virasorofield}) at central charge $c^\bullet=0$.
\end{example}
\begin{example}\label{topo}
[\textsl{topological $N=2$ superconformal algebra}]\hspace*{\fill}

\noindent
With $\mathcal A_D$, $\mathbb H$, and the fields of the $bc-\beta\gamma$-system defined
in the above 
Example \ref{bcbetagamma}, let
\begin{equation}\label{topN2}
J(z)  \;:=\; \sum_{j=1}^D :\!\varphi^j\psi^j\!\!:\!\!(z), \;\;
Q(z) \;:=\;
\sum_{j=1}^D :\! a^j\varphi^j\!\!:\!\!(z), \;\;
G(z) \;:=\; \sum_{j=1}^D :\!\psi^j\partial b^j\!\!:\!\!(z).
\end{equation}
These fields obey the so-called \textsc{topological $N=2$ superconformal algebra}
at central charge $c=3D$:
\begin{eqnarray}
T^{\rm top}(z)T^{\rm top}(w) &\sim&   {2T^{\rm top}(w)\over(z-w)^2}+ {\partial T^{\rm top}(w)\over z-w},
\label{topN=2a}\\
T^{\rm top}(z)J(w) &\sim&   -{c/3\over (z-w)^3}+{J(w)\over(z-w)^2}+ {\partial J(w)\over z-w}, 
\quad
J(z)J(w) \sim {c/3\over (z-w)^2}, \nonumber\\
T^{\rm top}(z)Q(w) &\sim&   {Q(w)\over(z-w)^2}+ {\partial Q(w)\over z-w}, \quad
Q(z)Q(w) \sim 0, 
\quad
J(z)Q(w) \sim {Q(w)\over z-w},\nonumber\\
T^{\rm top}(z)G(w) &\sim&   {2G(w)\over(z-w)^2}+ {\partial G(w)\over z-w}, 
\quad
G(z)G(w) \sim 0, 
\quad
J(z)G(w) \sim - {G(w)\over z-w},\nonumber
\\
Q(z)G(w) &\sim&   {c/3\over (z-w)^3}+{J(w)\over(z-w)^2}+ {T^{\rm top}(w)\over z-w}.
\label{topN=2o}
\end{eqnarray}
\end{example}
\begin{example}{}\label{N2example}
[\textsl{$N=2$ superconformal algebra}]\hspace*{\fill}

\noindent
Consider a $\mathbb C$-vectorspace $\mathbb H$ and pairwise local fields
$T^{\rm top}(z),\, J(z),\, Q(z),\, G(z)$ on $\mathbb H$ which obey the topological $N=2$ superconformal
algebra (\ref{topN=2a})--(\ref{topN=2o}) at central charge $c$. Now let
\begin{equation}\label{toptwist}
T(z):= T^{\rm top}(z) - {1\over2}\partial J(z),\quad
G^+(z):=Q(z),\quad G^-(z):=G(z).
\end{equation}
Then $T(z)$ is another Virasoro field as in (\ref{Virasorofield}), but now with
central charge $c^\bullet=c$, and 
the fields $T(z),\, J(z),\, G^+(z),\, G^-(z)$ on $\mathbb H$ obey the so-called
\textsc{$N=2$ superconformal algebra} at central charge $c$,
\begin{eqnarray}
T(z)T(w) &\sim&    {c/2\over (z-w)^4}+ {2T(w)\over(z-w)^2}+ {\partial T(w)\over z-w},\label{N=2a}
\\
T(z)J(w) &\sim&   {J(w)\over(z-w)^2}+ {\partial J(w)\over z-w}, 
\quad
J(z)J(w) \sim {c/3\over (z-w)^2}, \nonumber\\
T(z)G^\pm(w) &\sim&   {3/2G^\pm(w)\over(z-w)^2}+ {\partial G^\pm(w)\over z-w}, \quad
J(z)G^\pm(w) \sim \pm {G^\pm(w)\over z-w},\nonumber\\
G^\pm(z)G^\mp(w) &\sim&   {c/3\over (z-w)^3}\pm{J(w)\over(z-w)^2}+ {T(w)\pm{1\over2}\partial J(w)\over z-w},
\;\;\;
G^\pm(z)G^\pm(w)\; \sim\; 0.\nonumber\\[-1em]
\label{N=2o}
\end{eqnarray}
Equation (\ref{toptwist}) is referred to by the statement that the fields $T^{\rm top}(z),\, J(z),\, Q(z),\, G(z)$
are obtained from the fields $T(z),\, J(z),\, G^+(z),\, G^-(z)$ by a \textsc{topological A-twist}. Analogously, fields 
$T^{\rm top}(z),\, -J(z),\, Q(z),\, G(z)$ which obey a topological $N=2$ superconformal algebra at central charge $c$
are obtained from  fields $T(z),\, J(z)$, 
$G^+(z),\, G^-(z)$ which obey an $N=2$ superconformal algebra at central charge $c$
 by a \textsc{topological B-twist} iff $T^{\rm top}(z)=T(z) -{1\over2}\partial J(z),\, Q(z)=G^-(z),\, G=G^+(z)$,
 see \cite{wi88b,egya90}.
On the level of the $N=2$ superconformal algebras, the transition between topological A-twist and topological B-twist 
is induced by $(T,\, J,\, G^+,\, G^-)\mapsto (T,\, -J,\, G^-,\, G^+)$, an automorphism of the superconformal 
algebra. This automorphism is at the heart of \textsc{mirror symmetry} \cite{lvw89}.
\end{example}
We are now ready to define one of the fundamental 
ingredients of CFT, namely  the notion of \textsc{conformal vertex algebra}.
The 
definition is taken from \cite{frbe04} and follows \cite{fkrw95} and \cite{bo86,ka96}:
\begin{svgraybox}\vspace*{-1.5em}
\begin{definition}\label{voa}
A \textsc{conformal vertex algebra at central
charge $c\in\mathbb C$}
is given by the following data:
\begin{itemize}
\item
a $\mathbb Z$-graded
$\mathbb C$-vectorspace $W=\oplus_{m\in\mathbb Z}W_m$ 
called the \textsc{space of states},\index{state}
\item
a special vector $\Omega\in W_0$ called the \textsc{vacuum},
\item
a linear operator $L\colon W\rightarrow W$ 
called the \textsc{translation operator},
\item 
a special vector $T\in W_2$
called the \textsc{conformal vector},
\item
a linear map
$$
Y(\cdot,z)\colon W\longrightarrow\mbox{End}(W)[\hspace{-.1em}[z^{\pm1}]\hspace{-.1em}],
$$
called the \textsc{state-field correspondence}, which
assigns to every  $A\in W$ a field $A(z):=Y(A,z)$ on $W$.
\end{itemize}
These data obey the following axioms:
\begin{itemize}
\item
The \textsc{vacuum axiom}. We have $\Omega(z)=1$, 
and for every $A\in W$ and $A(z)=\sum_n\widehat A_n z^n$, we obtain
$A(z)\Omega\in W[\hspace{-.1em}[ z]\hspace{-.1em}]$, 
such that $A(z)\Omega$ is well-defined in $z=0$ and
$$
A(z)\Omega_{\mid z=0} 
=\widehat A_0\Omega
=A\in W.
$$
One says: the field $A(z)$ \textsc{creates} the state $A$ from the vacuum.
\item
The \textsc{translation axiom}. We have
$$
L\Omega=0 \quad\mbox{ and }\quad
\forall A\in W\colon\quad
\left[ L, A(z) \right] = \partial A(z).
$$
\item
The \textsc{locality axiom}. 

All fields
$A(z)$ with $A\in W$ are local with respect to each other.
\end{itemize}
The (ungraded) vectorspace $W$ with $\Omega$, $L$, and the map $Y$ is called a 
\textsc{vertex algebra}. In a \textsc{conformal} vertex algebra, in addition
\begin{itemize}
\item
The field $T(z)=\sum_{n=-\infty}^\infty L_n z^{n-2}$
associated to the conformal vector $T$ 
by the state-field correspondence is a Virasoro field
obeying the OPE (\ref{Virasorofield}) with central charge $c^\bullet=c$.
\item
The translation operator $L$ is given by $L=L_1$ and has  degree $1$.
\item
For all $m\in\mathbb Z$, 
$L_0{}_{\mid W_m}=m$, and
for $A\in W_m$, the field $A(z)$ has weight $m$, i.e.\
$A(z)=\sum_{n=-\infty}^\infty A_n z^{n-m}$ with $A_n\in\mbox{End}(W)$ of degree $n$.
\end{itemize}
\end{definition}
\vspace*{-1em}
\end{svgraybox}
In the context of superconformal field theories, the notion of conformal vertex algebras of 
Definition \ref{voa} is generalized to superconformal vertex algebras.
For an $N=2$ superconformal vertex algebra, 
the vectorspace $W$ in the above Definition is graded by ${1\over2}\mathbb Z$ instead of $\mathbb Z$, one needs to allow
odd fields $A(z)=Y(A,z)$, which can have mode expansions in 
$z^{1/2}\cdot\mbox{End}(W)[\hspace{-.1em}[z^{\pm1}]\hspace{-.1em}]$,
 and one needs to generalize the notion of locality to such fields,
as explained in the discussion of Definition \ref{locality}. Finally,  one needs to assume that
there exist special states $J\in W_1$ and $G^\pm\in W_{3/2}$ such that the associated 
fields $J(z),\,G^\pm(z)$
obey the $N=2$ superconformal algebra (\ref{N=2a})--(\ref{N=2o}). \\

An important ingredient of CFT are the so-called $n$-point functions, which associate a function in 
$n$ complex variables to every $n$-tuple of states in the CFT. 
These $n$-point functions are naturally related to the notion of vertex algebras, as we shall
illustrate now.
Assume that $W$ is the $\mathbb Z$-graded vectorspace which underlies a 
conformal vertex algebra, with notations as in Definition \ref{voa}. Furthermore, assume
that $W$ comes equipped with a positive definite scalar product $\langle\cdot,\cdot\rangle$, such that 
$W=\oplus_{m\in\mathbb Z} W_m$ is an orthogonal direct sum. 
Let $A(z),\,B(w)$ denote the fields associated to  $A,\,B\in W$ by the state-field correspondence,
which are local with respect to each other by the locality axiom.
Hence by the very Definition \ref{locality} of locality, the formal power series
$\langle\Omega, A(z)B(w)\Omega\rangle$ and $\langle\Omega, B(w)A(z)\Omega\rangle$ 
are obtained from the same series in $\mathbb C[\hspace{-.1em}[ z, w]\hspace{-.1em}][z^{-1}, w^{-1}, (z-w)^{-1}]$
by means of the (formal) Taylor expansions $\iota_{z>w}$ and $\iota_{w>z}$, respectively. 
This series is denoted by 
$\langle A(z)B(w)\rangle\in \mathbb C[\hspace{-.1em}[ z, w]\hspace{-.1em}][z^{-1}, w^{-1}, (z-w)^{-1}]$,
such that 
$$
\iota_{z>w}\left( \langle A(z)B(w)\rangle \right) = \langle\Omega, A(z)B(w)\Omega\rangle,\quad
\iota_{w>z}\left( \langle A(z)B(w)\rangle \right) = \langle\Omega, B(w)A(z)\Omega\rangle.
$$
Then $\langle A(z)B(w)\rangle$ is an example of a $2$-point function, and for $A_1,\,\ldots,\,A_n\in W$ one 
analogously defines the $n$-point functions $\langle A_1(z_1)\cdots A_n(z_n)\rangle$ by successive
OPE. The additional properties of CFTs ensure that these $n$-point functions define meromorphic 
functions in complex variables $z_1,\,\ldots, z_n\in\mathbb C$, whose possible poles 
are restricted to the partial diagonals
$z_i=z_j,\, i\neq j$. 
\subsection{Defining conformal field theories}\label{cftdef}
This section summarizes an axiomatic approach to conformal field theory. Instead of  a full account, the 
focus lies on those ingredients of CFTs that are relevant for the remaining sections 
of this exposition. More details can be found e.g.\ in \cite{we04,we10}.
We list the ingredients and defining properties of a two-dimensional Euclidean unitary conformal field theory
at central charges $c,\,\overline c$:
\newcounter{kwingredient}\setcounter{kwingredient}{0}\renewcommand{\thekwingredient}{\Roman{kwingredient}}
\newcounter{kwproperty}\setcounter{kwproperty}{0}\renewcommand{\thekwproperty}{\Alph{kwproperty}}
\begin{svgraybox}\vspace*{-1.5em}
\runinhead{Ingredient \Roman{kwingredient}.}\refstepcounter{kwingredient}\label{statespace}
[The  \textsc{space of states} $\mathbb H$]

\noindent
The space $\mathbb H$ is 
a $\mathbb C$-vectorspace with positive definite scalar product
$\langle\cdot,\cdot\rangle$ and with a compatible real structure $v\mapsto v^\ast$. 
Furthermore, there are 
two Virasoro fields $T(z),\, \overline T(\overline z)$ of central charges $c,\,\overline c$
on $\mathbb H$, see equation (\ref{Virasorofield}), where the OPE between $T$ and $\overline T$ is
trivial:
$$
T(z)\overline T(\overline z) \sim 0. 
$$
\vspace*{-2em}
\end{svgraybox}
\noindent
The space  of states of a CFT  must have a number of additional 
 properties:
\begin{svgraybox}\vspace*{-1.5em}
\runinhead{Property \Alph{kwproperty}.}\refstepcounter{kwproperty}\label{unitarity}
The space of states $\mathbb H$ furnishes a \textsc{unitary} representation of the
two commuting copies of a Virasoro algebra generated by the
modes $L_n,\,\overline L_n,\;n\in\mathbb Z$, of the Virasoro fields $T(z)$ and $\overline T(\overline z)$,
which is \textsc{compatible with the real structure} of $\mathbb H$. The central elements
$c,\,\overline c$ act by multiplication with fixed, real constants, also denoted $c,\,\overline c\in\mathbb R$.
The operators $L_0$ and $\overline L_0$ are self-adjoint and positive semidefinite, and $\mathbb H$
decomposes into a direct sum of their simultaneous eigenspaces indexed by
$R\subset\mathbb R^2$,
$$
\mathbb H = \bigoplus_{(h,\overline h)\in R} \mathbb H_{h,\overline h}, \quad
\mathbb H_{h,\overline h}:=\mbox{ker}\left( L_0-h\cdot\mbox{id}\right)\cap \mbox{ker}\left( \overline L_0-\overline h\cdot\mbox{id}\right).
$$
By this we mean that $R$ does not have accumulation points, and that
every vector in $\mathbb H$ is a sum of contributions from finitely many different
eigenspaces $\mathbb H_{h,\overline h}$. Moreover,  every
$\mathbb H_{h,\overline h}$ is finite dimensional.\\
\vspace*{-1.5em}\end{svgraybox}

Property \ref{unitarity}  ensures that the space of states $\mathbb H$ of every
conformal field theory furnishes a very well-behaved representation of two commuting
copies of a Virasoro algebra. In addition, we need to assume that the \textsc{character}
of this representation has favorable properties:
\begin{svgraybox}\vspace*{-1.5em}
\runinhead{Property \Alph{kwproperty}.}\refstepcounter{kwproperty}\label{partfun}
For $\tau\in\mathbb C,\,\Im(\tau)>0$, let $q:=\exp(2\pi i\tau)$; the \textsc{partition function}
$$
Z(\tau) 
:= \sum_{(h,\overline h)\in R} 
\left(\mbox{dim}_{\mathbb C} {\mathbb H}_{h,\overline h}\right) 
q^{h-c/24} \overline q^{\overline h-\overline c/24}
= \mbox{Tr}_{\mathbb H} \left(q^{L_0-c/24} \overline q^{\overline L_0-\overline c/24}\right)
$$
is well defined for all
values of $\tau$ in the complex upper halfplane, and it is invariant under
modular transformations
$$
\tau\mapsto \frac{a\tau+b}{c\tau+d}\,,\quad 
\left(\begin{array}{cc}a&b\\c&d\end{array}\right)\in SL(2,\mathbb Z).
$$
\vspace*{-2em}\end{svgraybox}
Since by Property \ref{partfun} the partition function is modular invariant, it in particular is
invariant under the translation $\tau\mapsto\tau+1$ of the modular parameter. This implies
that for every pair $(h,\overline h)\in R$ of eigenvalues of $L_0$ and $\overline L_0$,
we have $h-\overline h\in\mathbb Z$. Hence the subspaces 
$W:=\mbox{ker}\left( \overline L_0\right)$ and $\overline W:=\mbox{ker}\left(  L_0\right)$
are $\mathbb Z$-graded by $L_0$ and $\overline L_0$, respectively. To obtain a CFT,
these subspaces are required to carry additional structure, which we are already familiar with:
\begin{svgraybox}\vspace*{-1.5em}
\runinhead{Property \Alph{kwproperty}.}\refstepcounter{kwproperty}\label{chiral}
The subspaces $W:=\mbox{ker}\left( \overline L_0\right)$ and $\overline W:=\mbox{ker}\left(  L_0\right)$
of $\mathbb H$ carry the structure of conformal vertex algebras, see Definition \ref{voa}, with $T(z)$
and $\overline T(\overline z)$ the fields associated to the respective conformal vectors
by the state-field correspondence. Moreover,
the vacuum vector $\Omega$ of the conformal vertex algebra $W$ agrees with the
vacuum vector of $\overline W$, and $\Omega$ is a real unit vector yielding a basis of 
$W\cap\overline W=\mathbb H_{0,0}$.\\
The vertex algebras with underlying vectorspaces
$W$ and $\overline W$ are called the \textsc{chiral algebras} of the CFT,
and to simplify the terminology, we also refer to $W$ and $\overline W$
as the chiral algebras. 
\vspace*{-0.5em}\end{svgraybox}
As was discussed at the end of Section \ref{vertexalgebras}, in this setting there is a 
natural definition of $n$-point functions for the fields in the chiral algebras
associated to 
$W$ and $\overline W$. This definition, however, is not sufficient to capture the general $n$-point functions
of conformal field theory. The notion is generalized along the following lines:
\begin{svgraybox}\vspace*{-1.5em}
\runinhead{Ingredient \Roman{kwingredient}.}\refstepcounter{kwingredient}\label{npoints}
[The system $\langle\cdots\rangle$ of $n$-point functions]

\noindent
The space of states $\mathbb H$ is equipped with a system $\langle\cdots\rangle$ of 
\textsc{$n$-point functions},
that is, for every $n\in\mathbb N$ we have  a map
$$
\mathbb H^{\otimes n}\longrightarrow \mbox{Maps}(\mathbb C^n\setminus\bigcup_{i\neq j}D_{i,j},\mathbb C),
\quad
D_{i,j} :=\left\{ (z_1,\,\ldots,\,z_n)\in\mathbb C^n\mid z_i= z_j\right\},
$$
which is compatible with complex conjugation, 
and such that every function in the image is real 
analytic and allows
an appropriate expansion about every partial diagonal $D_{i,j}$.
\vspace*{-0.5em}\end{svgraybox}
The following Property \ref{localnpts}, which along with Property \ref{Poincare} 
governs the behavior of  the $n$-point functions, is immediate 
on the chiral algebras $W$ and $\overline W$,
by definition:
\begin{svgraybox}\vspace*{-1.5em}
\runinhead{Property \Alph{kwproperty}.}\refstepcounter{kwproperty}\label{localnpts}
The $n$-point functions are  \textsc{local}, that is, 
for every permutation $\sigma\in S_n$ and all $\phi_i\in\mathbb H$,
$$
\langle \phi_1(z_1)\cdots\phi_n(z_n) \rangle
= \langle \phi_{\sigma(1)}(z_{\sigma(1)})\cdots
\phi_{\sigma(n)}(z_{\sigma(n)}) \rangle.
$$
\vspace*{-2em}\end{svgraybox}
Consider an $n$-point function $\langle \phi(z_1)\cdots\phi(z_n) \rangle$ 
with $\phi=\phi_1=\cdots=\phi_n\in\mathbb H$ as a
function of one complex variable $z=z_k$, while all other $z_l,\, l\neq k,$ are fixed. 
The closure of the domain of definition of this function is  the \textsc{worldsheet} on which the 
$n$-point function is defined. Therefore,  Ingredient \ref{npoints} 
yields $n$-point functions whose worldsheet is the Riemann sphere $\overline{\mathbb C}$.
As a basic feature of conformal field theory, the $n$-point functions are assumed to 
transform covariantly under conformal maps between worldsheets. In particular,
\begin{svgraybox}\vspace*{-1.5em}
\runinhead{Property \Alph{kwproperty}.}\refstepcounter{kwproperty}\label{Poincare}
The $n$-point functions are \textsc{Poincar\'e covariant}, that is,
 for all isometries
and all dilations $f$ of the Euclidean plane $\mathbb C$, and 
for all $\phi_i\in\mathbb H_{h_i,\overline h_i}$,
$$
\langle \phi_1(f(z_1))\cdots\phi_n(f(z_n)) \rangle
= 
\prod_{i=1}^n 
\left[\left( f^\prime(z_i) \right)^{-h_i}
\left( \smash{\overline{f^\prime(z_i)}} \right)^{-\overline h_i}\right]
\langle \phi_1(z_1)\cdots\phi_n(z_n) \rangle,
$$
where $f^\prime(z)=\partial_z f(z)$. Moreover,
\textsc{infinitesimal translations}
$\alpha L_1+\overline\alpha\overline L_1,\,\alpha, \overline\alpha\in\mathbb C$,
are represented by 
$\alpha\partial_z+\overline\alpha\partial_{\overline z}$, i.e.\
for arbitrary $\phi_i\in\mathbb H$,
\begin{eqnarray*}
\langle \phi_1(z_1)\cdots\phi_{n-1}(z_{n-1})(L_1 \phi_n)(z_n) \rangle
&=& {\partial\over\partial z_n}
\langle \phi_1(z_1) \cdots\phi_{n-1}(z_{n-1}) \phi_n(z_n) \rangle,\\
\langle \phi_1(z_1)\cdots\phi_{n-1}(z_{n-1})(\overline L_1 \phi_n)(z_n) \rangle
&=& {\partial\over\partial\overline z_n}
\langle \phi_1(z_1) \cdots\phi_{n-1}(z_{n-1}) \phi_n(z_n) \rangle.
\end{eqnarray*}
\vspace*{-2em}\end{svgraybox}
The remaining requirements on the $n$-point functions, unfortunately, are rather more
involved. Roughly, they firstly generalize Property \ref{Poincare} by 
ensuring that the representation of the two commuting copies of the Virasoro algebra 
on $\mathbb H$ (see Property \ref{unitarity})
induces an action by infinitesimal conformal transformations on the worldsheet.  
Furthermore, the operator product expansion 
of Definition \ref{locality} is generalized to induce the appropriate expansions
of the $n$-point functions about partial diagonals, see Ingredient \ref{npoints}.
Finally, $n$-point functions must be defined on worldsheets
with arbitrary genus.  
Since these additional properties are not needed explicitly in the remaining sections of the present exposition, 
here only the relevant keywords are listed in the final
\begin{svgraybox}\vspace*{-1.5em}
\runinhead{Property \Alph{kwproperty}.}\refstepcounter{kwproperty}\label{finalprop}
The system $\langle\cdots\rangle$ of $n$-point functions
is \textsc{conformally covariant}, and it \textsc{represents
an operator product expansion} such that \textsc{reflection 
positivity} holds. Moreover, the \textsc{universality condition} holds, and
sewing allows to define $n$-point functions on \textsc{worldsheets of arbitrary genus}.
\vspace*{-0.5em}\end{svgraybox}
As was mentioned at the beginning of this section, the  ingredients of CFTs listed above 
yield \textsl{two-dimensional Euclidean unitary conformal 
field theories}. Indeed, these adjectives have been implemented in Properties \ref{unitarity}--\ref{finalprop}:
according to the discussion that precedes Property \ref{Poincare} along with Property \ref{finalprop},
the worldsheets of our CFTs are 
\textsl{two-dimensional  Euclidean} manifolds.
\textsl{Conformality} is implemented by means of the two commuting copies of the Virasoro algebra,
see the discussion of equations (\ref{Virasorofield}) and (\ref{Virasoroalgebra}), which act
by infinitesimal conformal transformations on the worldsheets of the $n$-point functions 
by Properties \ref{Poincare} and \ref{finalprop}. On 
 the space of states $\mathbb H$, Property \ref{unitarity} ensures that the representation
of the infinitesimal conformal transformations is \textsl{unitary}.
\\

Our approach to CFT is convenient, since it concretely implements the interplay between representation
theory with the analytic properties of the $n$-point functions, which is characteristic of two-dimensional
conformal quantum field theories. However, the relation to  more general 
quantum field theories (QFTs) is not so evident. Let us briefly comment on this connection. 

First, for the relevant QFTs
we restrict to \textsl{Euclidean} quantum field theories according to a system of axioms that are based 
on the \textsc{Osterwalder-Schrader  axioms} \cite{ossc73,ossc75}, see  \cite{ffk89,sch97}.
According to \cite{ossc73,ossc75}, these axioms ensure that from such a QFT one
can construct a Hilbert space $\widetilde{\mathbb H}$ of states $\phi$ and associated fields $Y_\phi$, 
where each field $Y_\phi$ yields a 
densely defined  linear operator $Y_\phi(h)$ on $\widetilde{\mathbb H}$ for every test function $h$.
Moreover, there is a special state $\Omega$ 
which plays the role of the vacuum  as in our Property \ref{chiral}.

The Osterwalder-Schrader axioms require the existence of correlation functions associated to every
$n$-tuple of states in $\widetilde{\mathbb H}$ which resemble the $n$-point functions of CFT 
according to our Ingredient \ref{npoints}. 
To obtain the fields of CFT from those of the general QFT, 
one needs to perform a procedure called \textsc{localization}. 
Within the Hilbert space $\widetilde{\mathbb H}$
one  restricts to the subspace $\mathbb H$ which is generated by those states that are created by
the localized field operators from the vacuum, generalizing the vacuum 
axiom of our Definition \ref{voa}.
The Osterwalder-Schrader axioms then ensure that locality (Property \ref{localnpts}), Poincar\'e covariance under
isometries
(Property \ref{Poincare}) and reflection positivity (Property \ref{finalprop}) hold for the $n$-point functions obtained
from the correlation functions of the QFT.
 
According to \cite{ffk89}, conformal covariance can be implemented by means of three additional
axioms, ensuring the covariance of the $n$-point functions under dilations (Property \ref{Poincare}), the existence
of the Virasoro fields (Ingredient \ref{statespace}) and  of an OPE (Properties \ref{chiral} and \ref{finalprop}) 
with all the 
necessary features. See \cite[\S9.3]{sch97} for an excellent account.

If a CFT is obtained from a conformally covariant QFT by localization, then one often says that the CFT
is the  \textsc{short distance 
limit} of the QFT. For details on this 
mathematical procedure see \cite{ffk89,frga93,was95}. To the author's know\-ledge, it is
unknown whether a CFT in the sense of our approach
can always be viewed as a
short distance limit of a full-fledged QFT.\\

With the above, we  do not claim to provide a 
\textsl{minimal} axiomatic approach to CFT. For example, the requirement of Property \ref{finalprop}
that $n$-point functions are well-defined on worldsheets of arbitrary genus implies modular invariance
of the partition function, which was assumed separately in Property \ref{partfun}. Indeed, the partition function 
$Z(\tau)$ is the $0$-point function on a worldsheet torus with modulus $\tau$, where conformal invariance 
implies that $Z(\tau)$ indeed solely depends on the complex structure represented by $\tau\in\mathbb C$, $\Im(\tau)>0$.
Property \ref{partfun} is stated separately  for clarity, and because modular invariance plays a crucial
role in the discussion of the elliptic genus in Section \ref{ellgen} which is also essential for the remaining
sections of this exposition, while we refrain from a detailed discussion of Property \ref{finalprop}.

Mathematical implications of  modular invariance for CFTs
were first pointed out by Cardy \cite{ca86}. He observed that for those theories that had been studied 
by Belavin, Polyakov and Zamolodchikov in their seminal paper \cite{bpz84}, and that describe  physical
phenomena in statistical physics, modular invariance of the partition function poses constraints on the 
operator content. These constraints can be useful for the classification of CFTs.

In special cases, modular invariance can be proven from first principles, assuming only that the
$n$-point functions are well-defined on the Riemann sphere. In   \cite{na91}, Nahm argues that the assumption
that the $n$-point functions on the torus define thermal states of the field algebra, which in turn is of type I, suffices
to deduce modular invariance. Under an assumption  known as \textsc{Condition $C$} or   \textsc{Condition $C_2$}, 
which amounts to certain quotients of the chiral algebras being finite dimensional, 
Zhu proves in \cite{zh96} that modular invariance follows, as well. This covers a large class of examples of CFTs,
among them the ones studied by Belavin, Polyakov and Zamolodchikov.
\\

An \textsc{$N=(2,2)$ superconformal field theory}
is a CFT as above, where the notion of locality is generalized according to what was
said in the discussion of Definition \ref{locality}, and the representations of the
two commuting copies of a Virasoro algebra are extended to representations of
$N=2$ superconformal algebras, see equations (\ref{N=2a})--(\ref{N=2o}). As a first
additional ingredient to these theories one therefore needs
\begin{svgraybox}\vspace*{-1.5em}
\runinhead{Ingredient \Roman{kwingredient}.}\refstepcounter{kwingredient}\label{Z2grading}
[Compatible $\mathbb Z_2$-grading of the space of states]

\noindent
The space of states $\mathbb H$ carries a 
$\mathbb Z_2$-grading  $\mathbb H=\mathbb H_b\oplus\mathbb H_f$ into \textsc{worldsheet bosons}
$\mathbb H_b$ (even) and \textsc{worldsheet fermions} $\mathbb H_f$
(odd), which is compatible with Properties \ref{unitarity}--\ref{finalprop}.
\\
In more detail, for compatibility with Property \ref{unitarity},  the decomposition $\mathbb H=\mathbb H_b\oplus\mathbb H_f$ 
must be orthogonal and invariant under the action of the 
two commuting copies of the Virasoro algebra. In Property \ref{partfun}, the trace defining the  partition function   is 
taken over the bosonic subspace $\mathbb H_b$, only. The chiral algebras introduced in Property \ref{chiral}
must contain $N=2$ superconformal vertex algebras as introduced in the discussion of Definition \ref{voa},
whose modes act unitarily on $\mathbb H$. The
notion of locality in Property \ref{localnpts} is generalized to \textsc{semi-locality}, meaning that 
$$
\langle \phi_1(z_1)\cdots\phi_n(z_n) \rangle
= (-1)^I \langle \phi_{\sigma(1)}(z_{\sigma(1)})\cdots
\phi_{\sigma(n)}(z_{\sigma(n)}) \rangle
$$
if $\sigma\in S_n$ and all $\phi_i\in\mathbb H$ have definite parity. Here, $I$ is the number
of inversions of odd states in $\sigma$, that is, the number
of pairs $(i,j)$ of indices with $i<j$ and $\sigma(i)>\sigma(j)$ and such that 
$\phi_i,\,\phi_j\in \mathbb H_f$. Properties \ref{Poincare} and \ref{finalprop} remain 
unchanged.
\vspace*{-1.5em}\end{svgraybox}
The fields in the chiral algebras of the CFT that  furnish  the
two commuting copies of $N=2$ superconformal vertex algebras
according to Property \ref{Z2grading}
are 
generally denoted $T(z),\,J(z),\,G^+(z),\,G^-(z)$ and 
$\overline T(\overline z),\,\overline J(\overline z),\,\overline G^+(\overline z),\,\overline G^-(\overline z)$
with OPEs as in (\ref{N=2a})--(\ref{N=2o}).
The mode expansions for the even fields are denoted as 
\begin{equation}\label{modeexpansions}
T(z)=\sum_n L_n z^{n-2},\;\;
J(z)=\sum_n J_n z^{n-1},\;\; 
\overline T(\overline z)=\sum_n \overline L_n \overline z^{n-2},\;\;
\overline J(\overline z)=\sum_n \overline J_n \overline z^{n-1},
\end{equation}
in accord with Definition \ref{voa}. As mentioned in the discussion after Definition \ref{voa},
the odd fields $G^\pm(z)$ can have mode expansions either in 
$\mbox{End}_{\mathbb C}(\mathbb H)[\hspace{-.1em}[z^{\pm1}]\hspace{-.1em}]$ or in
$z^{1/2}\cdot\mbox{End}_{\mathbb C}(\mathbb H)[\hspace{-.1em}[z^{\pm1}]\hspace{-.1em}]$,
and analogously for $\overline G^\pm(\overline z)$. This induces another  
$\mathbb Z_2\times\mathbb Z_2$ grading of the space of states $\mathbb H$,
\begin{equation}\label{NSR}
\mathbb H=\mathbb H^{NS,NS}\oplus\mathbb H^{R,R}\oplus\mathbb H^{NS,R}\oplus\mathbb H^{R,NS},
\end{equation}
where  
$$
G^\pm(z)\in\mbox{End}_{\mathbb C}(\mathbb H^{NS,\bullet})[\hspace{-.1em}[z^{\pm1}]\hspace{-.1em}],\quad 
\overline G^\pm(\overline z)\in\mbox{End}_{\mathbb C}(\mathbb H^{\bullet,NS})[\hspace{-.1em}[\overline z^{\pm1}]\hspace{-.1em}]
$$
in the so-called \textsc{Neveu-Schwarz-} or \textsc{NS-sector}, while
$$
G^\pm(z)\in z^{1/2}\cdot\mbox{End}_{\mathbb C}(\mathbb H^{R,\bullet})[\hspace{-.1em}[z^{\pm1}]\hspace{-.1em}],\quad
\overline G^\pm(\overline z)\in\overline z^{1/2}\cdot\mbox{End}_{\mathbb C}(\mathbb H^{\bullet,R})[\hspace{-.1em}[\overline z^{\pm1}]\hspace{-.1em}]
$$
in the so-called \textsc{Ramond-} or \textsc{R-sector}. That is, 
on $\mathbb H^{S,\overline S}$ the fields $G^\pm(z)$ and $\overline G^\pm(\overline z)$
have mode expansions according to the $S$ and the $\overline S$ sector, respectively, 
with $S,\overline S\in\{R,NS\}$.

In what
follows, we  restrict our attention to so-called 
\textsc{non--chiral $N=(2,2)$ superconformal field theories with space-time supersymmetry}:
\begin{svgraybox}\vspace*{-1.5em}
\runinhead{Ingredient \Roman{kwingredient}.}\refstepcounter{kwingredient}\label{stSUSY}
[Space-time supersymmetry]

\noindent
The space of states $\mathbb H$ carries another compatible
$\mathbb Z_2$-grading by means of the properties of the odd fields
$G^\pm(z)$ and $\overline G^\pm(\overline z)$ of the $N=2$ superconformal vertex algebra into
$$
\mathbb H=\mathbb H^{NS}\oplus\mathbb H^{R}.
$$
Here, the decomposition (\ref{NSR}) reduces to $\mathbb H^{NS}:=\mathbb H^{NS,NS}$
and $\mathbb H^{R}:=\mathbb H^{R,R}$, while the sectors $\mathbb H^{NS,R}$ and $\mathbb H^{R,NS}$ are trivial.\\
Moreover,  as representations of the two commuting $N=2$ superconformal algebras
of Ingredient \ref{Z2grading}, $\mathbb H^{NS}$ and $\mathbb H^{R}$ are related by
an isomorphism 
$\Theta\colon \mathbb H \rightarrow\mathbb H$
which interchanges 
$\mathbb H^{NS}$ and $\mathbb H^{R}$ and which obeys
\begin{equation}\label{spfl}
\begin{array}{rclrcl}
[L_0,\Theta] &=&{c\over24}\Theta -{1\over2}\Theta\circ J_0,\quad&
[J_0,\Theta] &=&-{c\over6}\Theta ,\quad\\[5pt]
[\overline L_0,\Theta] &=&{\overline c\over24}\Theta -{1\over2}\Theta\circ \overline J_0,\quad&
[\overline J_0,\Theta] &=&-{\overline c\over6}\Theta,
\end{array}
\end{equation}
where $L_0,\, J_0,\,\overline L_0,\, \overline J_0$ are the zero-modes of the fields 
$T(z),\,J(z),\,\overline T(\overline z),\,\overline J(\overline z)$ obtained from the mode
expansions (\ref{modeexpansions}).
The isomorphism $\Theta$ is induced by a field of the theory called \textsc{spectral flow}, and it is also
known as \textsc{space-time supersymmetry}.
\vspace*{-0.5em}\end{svgraybox}
The spectral flow induces an inner automorphism $X\mapsto X^\prime=\Theta X \Theta^{-1}$
on the $N=(2,2)$ superconformal algebra, with the following beautiful property:
our standard generators $\left( X_n\right)_{X\in\{L,J,G^\pm\},\,n\in\mathbb Z}$
in the Ramond sector are mapped to generators $X_r^\prime$ in the Neveu-Schwarz
sector yielding the modes of fields $T^\prime,\,J^\prime,\,G^{\pm\prime}$ that obey
the $N=2$ superconformal algebra (\ref{N=2a})--(\ref{N=2o}), and
analogously for the right-moving fields. As such, spectral flow implements 
an equivalence of representations of the $N=(2,2)$ superconformal algebra.

With the above notion of CFT, a number of examples are known, like minimal models,
both the bosonic \cite{bpz84,gko86} and the supersymmetric ones
\cite{dpz86,bfk86,zafa86,qi87}. In string theory,  so-called non-linear sigma model constructions
are believed to provide a map from certain manifolds to
CFTs. While this construction is well understood for the simplest manifolds, namely for tori,
the mathematical details  in general are far from 
understood. 
\subsection{Example: toroidal conformal field theories}\label{toroidalcfts}
For illustration and for later reference, this section very briefly presents the class
of so-called \textsc{toroidal conformal field theories}. These theories are characterized by
the existence of ``sufficiently many $U(1)$-currents'' as in Example \ref{U1current} of Section 
\ref{vertexalgebras}.\\

We say that the chiral algebra $W=\oplus_m W_m$ of a CFT (see Property \ref{chiral}, Section \ref{cftdef})
\textsc{contains a $\mathfrak u(1)^d$-current algebra}, if $W_1$ contains an orthogonal
system $\left(a_1^k\Omega,\;k\in\{1,\,\ldots,\,d\}\right)$ of states, which under the state-field 
correspondence of Definition \ref{voa} have associated fields 
$j^k(z), \;k\in\{1,\,\ldots,\,d\}$, obeying the OPEs
\begin{equation}\label{currentalgebra}
\forall k,\, l\in \{1,\,\ldots,\,d\}\colon\quad\quad
j^k(z) j^l(w) \sim {\delta^{k,l}\over (z-w)^2}.
\end{equation}
For (bosonic) CFTs we then have
\begin{svgraybox}\vspace*{-1.5em}
\begin{definition}\label{bostoroidaldef}
A conformal field theory at central charges $c,\,\overline c$ is called \textsc{toroidal},
if $c=\overline c=d$ with $d\in\mathbb N$, and if the chiral algebras $W=\oplus_m W_m$, 
$\overline W=\oplus_m \overline W_m$ 
of Property \ref{chiral}, Section \ref{cftdef}, each
contain a $\mathfrak u(1)^d$-current algebra. 
\end{definition}
\vspace*{-1em}
\end{svgraybox}
For our purposes, the \textsc{toroidal $N=(2,2)$ superconformal  field theories} are more relevant.
They are characterized by the fact that their bosonic sector 
with space of states $\mathbb H_b$ contains a toroidal CFT 
at central charges $2D,\, 2D$ in the sense of 
Definition \ref{bostoroidaldef}, and in addition, they contain $D$
left- and $D$ right-moving  so-called \textsc{Dirac fermions with coupled
spin structures}. 
By this we mean first of all  that the subspace $W_{1/2}\subset W$ of the vectorspace underlying the chiral algebra
contains an orthogonal
system $\left((\psi_\pm^k)_{1/2}\Omega,\; k\in\{1,\,\ldots,\,D\}\right)$ of states, which under the state-field 
correspondence of Definition \ref{voa} have associated (odd) fields 
$\psi_k^\pm(z), \; k\in\{1,\ldots,D\}$, obeying the OPEs
\begin{equation}\label{Dirac}
\forall k,\, l\in \{1,\,\ldots,\,D\}\colon\quad\quad
\psi^+_k(z) \psi^-_l(w) \sim {\delta^{k,l}\over z-w}, \quad
\psi^\pm_k(z) \psi^\pm_l(w)\sim 0,
\end{equation}
and analogously for the subspace $\overline W_{1/2}\subset \overline W$ of the 
vectorspace underlying the second chiral algebra
in Property \ref{chiral}.
In addition,  all $\psi_k^\pm(z)$ are represented by
formal power series in $\mbox{End}_{\mathbb C}(\mathbb H^{NS})[\hspace{-.1em}[z^{\pm1}]\hspace{-.1em}]$
on $\mathbb H^{NS}$,
while on $\mathbb H^{R}$, they are represented  in
$z^{1/2}\cdot\mbox{End}_{\mathbb C}(\mathbb H^{R})[\hspace{-.1em}[z^{\pm1}]\hspace{-.1em}]$,
and analogously for the $\overline\psi_k^\pm(\overline z)$.
One shows that such a system of $D$ left- and $D$ right-moving Dirac fermions yields
a well-defined CFT at central charges $D,\,D$ (see e.g.\ \cite[\S8.2]{gi88b}).
\begin{svgraybox}\vspace*{-1.5em}
\begin{definition}\label{susytoroidaldef}
An $N=(2,2)$ superconformal field theory at central charges $c,\,\overline c$ with space-time
supersymmetry is toroidal, if $c=\overline c=3D$ with $D\in\mathbb N$, and if this theory is the
tensor product of a toroidal conformal field theory at central charges $2D,\, 2D$
according to Definition \ref{bostoroidaldef},
and a system of $D$ left- and $D$ right-moving Dirac fermions with coupled
spin structures. Moreover, the fields $\psi_k^\pm(z)$, $k\in\{1,\,\ldots,\,D\}$,
in (\ref{Dirac}) yield the superpartners of the $U(1)$-currents $j^l(z)$, $l\in\{1,\,\ldots,\,2D\}$,
in (\ref{currentalgebra}), and analogously for the right-moving fields. By this we mean that 
for the fields $G^\pm(z),\,\overline G^\pm(\overline z)$ in the 
two commuting superconformal vertex algebras (\ref{N=2a})--(\ref{N=2o}),
and with notations as above,
we have 
$$
\begin{array}{rclrcl}
\forall k\in\{1,\,\ldots,\,D\}\colon\;\;
G^\pm_{-1/2}a_1^k\Omega &=& (\psi_\pm^k)_{1/2}\Omega,\;&
G^\pm_{-1/2}a_1^{k+D}\Omega &=& \mp i(\psi_\pm^k)_{1/2}\Omega,\quad\\[5pt]
\overline G^\pm_{-1/2}\overline a_1^k\Omega &=& (\overline\psi_\pm^k)_{1/2}\Omega,\;& 
\overline G^\pm_{-1/2}\overline a_1^{k+D}\Omega &=& \mp i(\overline\psi_\pm^k)_{1/2}\Omega.
\end{array}
$$
\end{definition}
\vspace*{-1.5em}
\end{svgraybox}
The toroidal conformal and superconformal field theories have been very well understood by string theorists 
since the mid eighties \cite{cent85,na86}, and these theories  have also been reformulated
in terms of the vertex algebras presented in Section \ref{vertexalgebras} \cite{ka96,kaor00,frbe04}.
This includes the interpretation of the toroidal conformal field theories as non-linear sigma models
on tori, their deformations, and thus the structure of the moduli space of toroidal CFTs:
\begin{svgraybox}\vspace*{-1.5em}
\begin{theorem}[\cite{na86}]\label{toroidalmodulispace}
The moduli space $\mathcal M^{\rm tor}_D$
of  toroidal $N=(2,2)$ superconformal field theories 
at central charges $c=\overline c=3D$ with $D\in\mathbb N$ is a quotient of a $4D^2$-dimensional
Grassmannian by an infinite discrete group,
\begin{eqnarray*}
\mathcal M^{\rm tor}_D=
\mbox{O}^+(2D,2D;\mathbb Z)\backslash \mathcal T^{2D,2D},\\
\mbox{ where }
\mathcal T^{2D,2D}&:=& \mbox{O}^+(2D,2D;\mathbb R)\slash \mbox{SO}(2D)\times \mbox{O}(2D).
\end{eqnarray*}
Here, if $pq\neq0$, then
$\mbox{O}^+(p,q;\mathbb R)$ denotes the group of those elements in $\mbox{O}(p,q;\mathbb R)=\mbox{O}(\mathbb R^{p,q})$
which preserve the orientation of maximal positive definite oriented subspaces in $\mathbb R^{p,q}$,
and if $p\equiv q\mbox{ mod } 8$, then
$\mbox{O}^+(p,q;\mathbb Z)=\mbox{O}^+(p,q;\mathbb R)\cap\mbox{O}(\mathbb Z^{p,q})$
with $\mathbb Z^{p,q}\subset \mathbb R^{p,q}$ the standard even unimodular lattice of signature $(p,q)$.
\end{theorem}
\vspace*{-1em}\end{svgraybox}
\subsection{The elliptic genus}\label{ellgen}
In this section, 
the conformal field theoretic elliptic genus  is introduced and compared to the complex
elliptic genus that is known to topologists and geometers. 
This and the following section are completely expository with more details and
proofs to be found in the literature as referenced.\\

Let us first
consider an $N=(2,2)$ superconformal field theory at central charges $c,\,\overline c$ with space-time supersymmetry
according to the Ingredients \ref{statespace}--\ref{stSUSY} of Section \ref{cftdef}. 
For the zero-modes $J_0,\,\overline J_0$ of
the fields $J(z),\,\overline J(\overline z)$ in the two commuting 
$N=2$ superconformal vertex algebras 
$T(z),\,J(z),\,G^+(z),\,G^-(z),\,
\overline T(\overline z),\,\overline J(\overline z),\,\overline G^+(\overline z),\,\overline G^-(\overline z)$
of Ingredient \ref{Z2grading} according to (\ref{modeexpansions}) one finds:
these linear operators are self-adjoint and simultaneously diagonalizable
on the space of states $\mathbb H=\mathbb H^{NS}\oplus\mathbb H^R$. 
By Ingredient \ref{stSUSY},  the corresponding operator of spectral flow induces an equivalence of
representations $\mathbb H^{NS}\cong\mathbb H^R$ of  two $N=(2,2)$ superconformal algebras. This turns out to
imply that the linear operator $J_0-\overline J_0$ has only integral eigenvalues, which are even on 
$\mathbb H_b$ and odd on $\mathbb H_f$, see e.g. \cite[\S3.1]{diss}. Hence 
$(-1)^{J_0-\overline J_0}$ is an involution which
yields the $\mathbb Z_2$-grading $\mathbb H =\mathbb H_b\oplus\mathbb H_f$ and an induced
$\mathbb Z_2$-grading on $\mathbb H^R$. Following  \cite{eoty89}, this allows the definition of a supercharacter of the
superconformal field theory,
analogous to the  partition function in Property \ref{partfun}:
\begin{svgraybox}\vspace*{-1.5em}
\begin{definition}\label{cftellgen}
Consider
an $N=(2,2)$ superconformal field theory at central charges
$c,\,\overline c$ with space-time supersymmetry. Set
$q:=\exp(2\pi i\tau)$ for $\tau\in\mathbb C,\,\Im(\tau)>0$, and 
$y:=\exp(2\pi i z)$ for $z\in\mathbb C$. Then
\begin{eqnarray*}
\mathcal E(\tau,z) 
&:=& \mbox{Str}_{\mathbb H^{R}}\left( y^{J_0} q^{L_0-c/24} \overline q^{\overline L_0-\overline c/24}\right)\\
&=& \mbox{Tr}_{\mathbb H^{R}}\left( (-1)^{J_0-\overline J_0}y^{J_0} q^{L_0-c/24} \overline q^{\overline L_0-\overline c/24}\right)
\end{eqnarray*}
is the \textsc{conformal field theoretic elliptic genus} of the theory.
\end{definition}
\vspace*{-1em}
\end{svgraybox}
Using known properties of the $N=2$ superconformal algebra 
and  of its irreducible unitary representations, 
one shows (see  \cite{akmw87,eoty89,dfya93,wi94} for the original
results and e.g.\ \cite[\S3.1]{diss} for a summary and proofs):
\begin{svgraybox}\vspace*{-1.5em}
\begin{proposition}\label{ellgenprop}
Consider 
the conformal field theoretic elliptic genus $\mathcal E(\tau,z)$
of an $N=(2,2)$ superconformal field theory
at central charges $c,\,\overline c$ with space-time supersymmetry. 

\noindent
Then $\mathcal E(\tau,z)$ is holomorphic in $\tau$ and bounded when $\tau\rightarrow i\infty$.

\noindent
It is invariant under smooth deformations of the 
underlying superconformal field theory to any other space-time supersymmetric $N=(2,2)$ 
superconformal field theory with the same central charges. 

\noindent
Moreover, $\mathcal E(\tau,z)$ transforms covariantly under modular
transformations, 
$$
\mathcal E(\tau+1,z)= \mathcal E(\tau,z), \quad 
\mathcal E(-1/\tau,z/\tau) = e^{2\pi i {c\over6}\cdot{z^2\over\tau}}\mathcal E(\tau,z).
$$
If in addition $c=\overline c\in3\mathbb N$, 
and all eigenvalues of $J_0$ and $\overline J_0$ in the Ramond sector 
lie in ${c\over6}+\mathbb Z$, then
$$
\mathcal E(\tau,z+1)= (-1)^{c\over3} \mathcal E(\tau,z), \quad 
\mathcal E(\tau, z+\tau) = q^{-{c\over6}} y^{-{c\over3}}  \mathcal E(\tau,z).
$$
In other words, $\mathcal E(\tau,z)$ is a \textsc{weak Jacobi form} 
\mbox{\rm(}with a character, if $c/3$ is odd\mbox{\rm)}
of \textsc{weight $0$}
and \textsc{index $c/6$}.
\end{proposition}
\vspace*{-1em}\end{svgraybox}
Note that the additional assumptions on the central charges and the eigenvalues of $J_0$
and $\overline J_0$ in the last statement of Proposition \ref{ellgenprop} are expected to hold
for superconformal field theories that are 
obtained by a non-linear sigma model construction from some 
 Calabi-Yau $D$-manifold
\cite{eoty89}.\\

On the other hand, following Hirzebruch's seminal work on multiplicative 
sequences and their genera \cite{hi66}, 
the elliptic genus is known to topologists as a 
ring homomorphism from the cobordism ring of smooth oriented compact manifolds 
into a ring of modular functions \cite{la88,hi88}. 
For simplicity we assume that our underlying
manifold $X$ is a 
 Calabi-Yau $D$-manifold. Then its associated complex elliptic genus $\mathcal E_X(\tau, z)$
can be viewed as a modular function obeying the transformation properties of Proposition \ref{ellgenprop}
with $c=3D$ and interpolating between the standard topological invariants of $X$, namely its
\textsc{Euler characteristic $\chi(X)$}, its \textsc{signature $\sigma(X)$}, and its 
\textsc{holomorphic Euler characteristic $\chi(\mathcal O_X)$}. 

To understand this in more detail,  first recall the definition of the topological invariants mentioned above:
for $y\in\mathbb C$  the Hirzebruch $\chi_y$-genus \cite{hi66} is defined by
$$
\chi_y(X):=\sum_{p,q=0}^D (-1)^q y^p h^{p,q}(X),
$$
where the $h^{p,q}(X)$ are the Hodge numbers of $X$. Then 
\begin{equation}\label{topinv}
\chi(X):=\chi_{-1}(X),\quad
\sigma(X):=\chi_{+1}(X),\quad
\chi(\mathcal O_X):=\chi_{0}(X).
\end{equation}
Note that by the usual symmetries among the Hodge numbers $h^{p,q}(X)$ of a complex
K\"ahler manifold $X$, the signature $\sigma(X)=\sum_{p,q} (-1)^{q} h^{p,q}(X)$
vanishes if the complex dimension $D$ of $X$ is odd;
we have thus trivially extended the usual definition of the signature on oriented compact manifolds
whose real dimension is divisible by $4$ to all compact complex K\"ahler manifolds.

To motivate a standard formula for the specific 
elliptic genus which is of relevance to us, see Definition \ref{geoellgen},
we draw the analogy to the interpretation of the topological invariants
(\ref{topinv}) in terms of the Atiyah-Singer Index Theorem \cite{atsi63}.
For any complex vector bundle $E$ on $X$ and a formal variable $x$, we introduce the shorthand notations
$$
\Lambda_x E := \bigoplus_p x^p \Lambda^p E,\quad
S_x E:= \bigoplus_p x^p S^p E,
$$
where $\Lambda^p E,\, S^p E$ denote the exterior and the symmetric powers of $E$, respectively,
along with the \textsc{Chern character} on such formal power series in $x$
whose coefficients are complex vector bundles $F_p$:
$$
\mbox{ch}(\bigoplus_p x^p F_p) := \sum_p x^p \mbox{ch}(F_p).
$$
Then by the Hirzebruch-Riemann-Roch formula \cite{hi54}, which can be viewed as a special case of the
Atiyah-Singer Index Theorem, one finds
\begin{equation}\label{ashrr}
\chi_y(X) = \int_X \mbox{Td}(X) \mbox{ch} (\Lambda_yT^\ast),
\end{equation}
where $\mbox{Td}(X)$ denotes the \textsc{Todd genus}  and
$T:=T^{1,0}X$ is the holomorphic tangent bundle of $X$. 
Generalizing the expression in equation (\ref{ashrr}) and following \cite{hi88,wi88,kr90} we now
have
\begin{svgraybox}\vspace*{-1.5em}
\begin{definition}\label{geoellgen}
Let $X$ denote a compact complex $D$-manifold with holomorphic tangent bundle $T:=T^{1,0}X$.
Set
$$
\mathbb E_{q,-y}
:= y^{-D/2}  \bigotimes_{n=1}^\infty \left( \Lambda_{-yq^{n-1}} T^\ast\otimes \Lambda_{-y^{-1}q^n} T
\otimes S_{q^n} T^\ast\otimes S_{q^n} T\right),
$$
viewed as a formal power series with variables $y^{\pm1/2},\,q$,
whose coefficients are holomorphic vector bundles on $X$.

Analogously to Definition \ref{fielddef}, 
the integral $\int_X$ is extended linearly to  the vectorspace of formal power series whose coefficients
are  characteristic classes on $X$. Then with $q:=\exp(2\pi i\tau)$ and 
$y:=\exp(2\pi i z)$, the holomorphic Euler characteristic of $\mathbb E_{q,-y}$,
$$
\mathcal E_X(\tau,z) := \int_X \mbox{Td}(X) \mbox{ch}( \mathbb E_{q,-y} ) \quad
\in\;\; y^{-D/2}\cdot\mathbb Z[\hspace{-.1em}[y^{\pm1}, q]\hspace{-.1em}],
$$
is the (complex) \textsc{elliptic genus} of $X$.
\end{definition}
\vspace*{-1em}
\end{svgraybox}
By \cite{hi88,wi88,kr90}, the elliptic genus $\mathcal E_X(\tau,z)$ 
in fact yields a well-defined function in  $\tau\in\mathbb C$ with $\Im(\tau)>0$ and in
 $z\in\mathbb C$. If $X$ is a Calabi-Yau $D$-manifold, then $\mathcal E_X(\tau,z)$
is a {weak 
Jacobi form} (with a character, if $D$ is odd)
of {weight} $0$ and {index} $D/2$
\cite{boli00}. In other words,
with $c:=3D$ the elliptic genus $\mathcal E_X(\tau,z)$
obeys the transformation properties stated for $\mathcal E(\tau,z)$
in Proposition \ref{ellgenprop}, and it is bounded when $\tau\rightarrow i\infty$. 
One  checks that by definition, the elliptic genus indeed 
is a topological invariant which interpolates
 between the standard topological invariants of equation (\ref{topinv}), namely
 $\mathcal E_X(\tau, z)\stackrel{\tau\rightarrow i\infty}{\longrightarrow} y^{-D/2}\chi_{-y}(X)$
 and
 \begin{equation}\label{ellgeninterpolates}
 \begin{array}{rcl}
 \mathcal E_X(\tau, z=0) \;=\; \chi(X),\quad
 \mathcal E_X(\tau, z=1/2) &=& (-1)^{D/2}\sigma(X) + \mathcal O(q),\quad\\[5pt]
 q^{D/4}\mathcal E_X(\tau, z=(\tau+1)/2) &=& (-1)^{D/2}\chi(\mathcal O_X)+ \mathcal O(q).\quad
 \end{array}
 \end{equation}
According to Witten \cite{wi87,wi88}, 
the expression for the elliptic genus $\mathcal E_X(\tau,z)$ in Definition \ref{geoellgen}
can be interpreted as a regularized version of a $U(1)$-equivariant
index of a Dirac-like operator on the loop space of $X$, see also \cite{la88}. 
This explains the notation chosen in  Definition \ref{geoellgen}, 
and it also motivates why one expects that  
for CFTs which are obtained by a non-linear sigma model construction
from some  Calabi-Yau $D$-manifold $X$,  
the conformal field theoretic elliptic genus of Definition \ref{cftellgen}
agrees with the
{complex elliptic genus} of $X$ as in Definition \ref{geoellgen}. 
Note that the resulting equation 
\begin{equation}\label{mckeantype}
%\int_X \mbox{Td}(X) \mbox{ch}( \mathbb E_{q,-y} )  = 
\mathcal E_X(\tau,z) =
\mbox{Str}_{\mathbb H^{R}}\left( y^{J_0} q^{L_0-c/24} \overline q^{\overline L_0-\overline c/24}\right)
\end{equation}
would furnish a natural generalization of the \textsc{McKean-Singer  Formula} \cite{mksi67}.
While non-linear sigma model
constructions are not understood sufficiently well to even attempt a
general proof of this equation,
there is some evidence for its truth. On the one hand, 
as was pointed out in Section \ref{toroidalcfts},
$N=(2,2)$ superconformal field theories obtained from a non-linear sigma model on a
complex torus are very well understood.
One confirms that  their conformal field 
theoretic elliptic genus vanishes, 
as does the complex elliptic genus of a complex torus. Equation 
(\ref{mckeantype}) is also compatible with the construction of symmetric powers
of the manifold $X$ \cite{dmvv97}.
Moreover, 
compatibility of the elliptic genus with orbifold constructions was proved in 
\cite{boli00a,frsz07}. Further evidence in favor of the expectation 
(\ref{mckeantype}) arises from a discussion of the
chiral de Rham complex, see Section \ref{chiderham}. 
\subsection{The chiral de Rham complex}\label{chiderham}
As was pointed out above, non-linear sigma model constructions of $N=(2,2)$ superconformal
field theories are in general not very well understood. Therefore, a direct proof of the expected
equality (\ref{mckeantype}) is out of reach. However, instead of a full-fledged superconformal field theory,
in \cite{msv98} the authors construct a \textsl{sheaf of superconformal vertex algebras}, known as the
\textsc{chiral de Rham complex} $\Omega_X^{\rm ch}$, on any complex manifold $X$. The chiral de Rham
complex of $X$ is expected to be closely related to the non-linear sigma model on $X$, as we
shall discuss in the present section. \\

Let us begin by summarizing the construction of the chiral de Rham complex $\Omega_X^{\rm ch}$ 
for a complex $D$-dimensional manifold $X$, see \cite{msv98,goma03,lili07,bhs08}. 
First, to any coordinate neighborhood $U\subset X$
with holomorphic coordinates $(z^1,\,\ldots,\,z^D)$ one associates a $bc-\beta\gamma$ system
$\Omega_X^{\rm ch}(U)$ as in Example \ref{bcbetagamma}, see Section \ref{vertexalgebras}. Here, the even fields
$a^j,\, b^j$ are interpreted as arising from quantizing the local sections
${\partial/\partial z^j},\, z^j$ of the  sheaf of polyvector fields on $X$, while
the odd fields $\phi^j,\, \psi^j$ correspond to the local sections 
$dz^j,\, \partial/\partial(dz^j)$ of the  sheaf of differential operators on the de Rham algebra of differential forms. 
Indeed, by (\ref{bcbgalgebra}) the map 
$$
({\partial/\partial z^j},\, z^j,\,dz^j,\, \partial/\partial(dz^j))\longmapsto (a^j_0,\, b^j_0,\,\phi^j_0,\, \psi^j_0)
$$
induces a super-Lie algebra homomorphism.

According to \cite{msv98}, coordinate transforms on $X$
induce corresponding transformation rules for the fields  $a^j,\, b^j,\,\phi^j,\, \psi^j$ which are compatible with
the structure of the $bc-\beta\gamma$-system as discussed in Example \ref{bcbetagamma}. This allows
to glue the $\Omega_X^{\rm ch}(U)$ accordingly, and by localization, one indeed obtains a well-defined
sheaf of  vertex algebras over $X$, with a (non-associative) action of $\mathcal O_X$ on it. 

A key result of \cite{msv98} is the fact that under appropriate assumptions on $X$,
there are well-defined global sections of the sheaf 
$\mbox{End}_{\mathbb C}(\Omega_X^{\rm ch})[\hspace{-.1em}[z^{\pm1}]\hspace{-.1em}]$,
which are locally given by the fields (\ref{topT}), (\ref{topN2})
of the topological $N=2$ superconformal algebra (\ref{topN=2a})--(\ref{topN=2o}) discussed
in Example \ref{topo} of Section \ref{vertexalgebras}:
\begin{svgraybox}\vspace*{-1.5em}
\begin{theorem}[\cite{msv98}]\label{chiraldeRhamcomplex}
Let $X$ denote a compact complex manifold of dimension $D$.
As discussed above,  there is an associated sheaf
$\Omega_X^{\rm ch}$  of  vertex algebras on $X$. On every holomorphic 
coordinate chart $U\subset X$,
let $T^{\rm top}(z),\, J(z),\, Q(z),\, G(z)$ denote the local sections in 
$\mbox{\rm End}_{\mathbb C}(\Omega_X^{\rm ch}(U))[\hspace{-.1em}[z^{\pm1}]\hspace{-.1em}]$ defined
by (\ref{topT}), (\ref{topN2}), with mode expansions
$$
T^{\rm top}(z)=\sum_n L^{\rm top}_n z^{n-2},\;\; 
J(z)=\sum_n J_n z^{n-1},\; \;
Q(z)=\sum_n Q_n z^{n-1},\; \;
G(z)=\sum_n G_n z^{n-2},
$$
respectively. Then the following holds:
\begin{enumerate}
\item
The linear operators $F:=J_0$ and $d_{\rm dR}^{\rm ch}:=-Q_0$ are globally well-defined.
Moreover, $F$ defines a $\mathbb Z$-grading on $\Omega_X^{\rm ch}$, 
while $(d_{\rm dR}^{\rm ch})^2=0$, such that
$$
\forall p\in\mathbb Z\colon\quad
\Omega^{{\rm ch},p}_X(U):=\left\{ \Phi\in \Omega_X^{\rm ch}(U) \mid F\Phi=p\Phi\right\}
$$
yields a complex $(\Omega^{{\rm ch},\bullet}_X,\,d_{\rm dR}^{\rm ch})$, which
is called the \textsc{chiral de Rham complex}. 
\item
The map $(z^j,\,dz^j)\mapsto (b^j_0,\,\phi^j_0)$ induces a quasi-isomorphism from the usual
de Rham complex to the chiral de Rham complex of $X$.
\item
The  fields $T^{\rm top}(z)$ given locally in \mbox{\rm(\ref{topT})} define a global 
field on the chiral de Rham complex, by which we mean a global
section of the sheaf $\mbox{\rm End}_{\mathbb C}(\Omega_X^{\rm ch})[\hspace{-.1em}[z^{\pm1}]\hspace{-.1em}]$.
The
chiral de Rham
complex therefore is bigraded by $F$ and $L_0^{\rm top}$. 
\item
If $X$ is a Calabi-Yau manifold, then the  fields $J(z),\, Q(z),\, G(z)$ given
locally in \mbox{\rm(\ref{topN2})} also
define global fields on the chiral de Rham complex.
\end{enumerate}
\end{theorem}
\vspace*{-1em}\end{svgraybox}
As mentioned above, the sheaf $\Omega_X^{\rm ch}$ is not quasi-coherent. However, it has a
filtration which is compatible with the bigrading of Theorem \ref{chiraldeRhamcomplex} and such
that the corresponding graded object yields a quasi-coherent sheaf isomorphic to 
(the sheaf of sections of)
$(-y)^{D/2}\mathbb E_{q,y}$ as in Definition \ref{geoellgen}. This is used extensively
in \cite{bo01,boli00} to study the \textsl{\u Cech resolution} of the
\textsl{sheaf cohomology} $H^\ast(X,\Omega_X^{ch})$. 
Note that this means classical \u Cech cohomology, ignoring the differential $d_{\rm dR}^{\rm ch}$
of the chiral de Rham complex.
The authors  of \cite{bo01,boli00} find:
\begin{svgraybox}\vspace*{-1.5em}
\begin{theorem}[\cite{bo01,boli00}]\label{voafromchdr}
Consider a  Calabi-Yau $D$-manifold $X$, and 
the sheaf cohomology $H^\ast(X,\Omega_X^{ch})$ 
 of its chiral de Rham complex $\Omega_X^{ch}$.
Equip it with the induced bigrading by the operators $F=J_0$ and $L_0^{\rm top}$ of 
Theorem \ref{chiraldeRhamcomplex} 
and the  $\mathbb Z_2$-grading by $(-1)^{F+q}$ on $H^q(X,\Omega_X^{ch})$.
Then $H^\ast(X,\Omega_X^{ch})$ carries  a natural structure of a 
superconformal vertex  algebra, containing a topological $N=2$ superconformal 
 algebra \mbox{\rm\cite[Prop.~3.7 and Def.~4.1]{bo01}}.
Moreover  \mbox{\rm\cite{boli00}}, the \textsc{graded Euler characteristic} of the chiral de Rham complex, that is, the
supertrace of the operator $y^{-D/2}\cdot(y^{J_0}q^{L_0^{\rm top}})$ on $H^\ast(X,\Omega_X^{ch})$,
yields the elliptic genus $\mathcal E_X(\tau,z)$ of Definition \ref{geoellgen}.
\end{theorem}
\vspace*{-1em}\end{svgraybox}
Thus Theorem \ref{voafromchdr} indicates
a possible relationship between the chiral de Rham complex $\Omega_X^{\rm ch}$
of a Calabi-Yau $D$-manifold $X$ and a non-linear sigma model on $X$, since it recovers the 
(geometrically defined!) elliptic genus $\mathcal E_X(\tau,z)$ by means of a supertrace which at least in
spirit agrees with the expression on the right hand side of equation (\ref{mckeantype}).
Note that $L_0^{\rm top}=L_0-{1\over2}J_0$ by (\ref{toptwist}) (using Definition \ref{fielddef}
and (\ref{modeexpansions})). Therefore,
using the fact that the elliptic genus is holomorphic, along with the spectral flow (\ref{spfl}),
the conformal field theoretic elliptic genus of Definition \ref{cftellgen} can be expressed as
\begin{eqnarray*}
\mathcal E(\tau,z) 
&=&\mbox{Str}_{\mathbb H^{R}}\left( y^{J_0} q^{L_0-c/24}\right)\\
&=&y^{-c/6} \mbox{Str}_{\mathbb H^{NS}}\left( (yq^{-1/2})^{J_0} q^{L_0} \right)
\;=\; y^{-c/6} \mbox{Str}_{\mathbb H^{NS}}\left( y^{J_0} q^{L_0^{\rm top}} \right).
\end{eqnarray*}
Hence recalling $c=3D$ for a non-linear sigma model on 
a  Calabi-Yau $D$-manifold, 
one is led to conjecture that one might be able to identify
an appropriate cohomology of 
$\mathbb H^{NS}$  with $H^\ast(X,\Omega_X^{ch})$.

The details of such an identification are still more subtle, however. Indeed, by 
construction, the chiral de Rham complex depends only on the complex structure of $X$,
while the non-linear sigma model, in addition, depends on the complexified K\"ahler
structure of $X$. It is therefore natural to expect the vertex algebra of Theorem \ref{voafromchdr} 
to yield a truncated version of the non-linear sigma model by means
of the topological twists mentioned in Example \ref{N2example} of Section \ref{vertexalgebras}.
Since the crucial bundle $\mathbb E_{q,-y}$ of Definition \ref{geoellgen} resembles an infinite-dimensional
Fock space, while the traditional topological A- and B-twists yield finite dimensional spaces of states,
the so-called \textsc{half-twisted sigma model} according to Witten \cite{wi91} is the most natural candidate. 
It still cannot yield the vertex algebra
of Theorem \ref{voafromchdr}, since   it depends both on the complex
and on the complexified K\"ahler structure of $X$. Moreover, the \u Cech resolution, which is implicit
in $H^\ast(X,\Omega_X^{ch})$, does not resemble the standard
features of non-linear sigma models on $X$. 
According to Kapustin, however,  an infinite volume limit
of Witten's half twisted sigma model on $X$ yields the cohomology of $\Omega_X^{\rm ch}$
by means of yet another resolution of the complex, the so-called
 \textsc{Dolbeault resolution} \cite{ka05} .
\section{Conformal field theory on K3}\label{k3cft}
As emphasized repeatedly, non-linear sigma model constructions are in general not well understood,
except for the toroidal conformal field theories presented in Section \ref{toroidalcfts}. 
Recall however that there are only two topologically distinct types of 
 Calabi-Yau $2$-manifolds,
namely the complex $2$-tori and the K3 surfaces (see e.g.\ \cite[Ch.VIII]{bhpv84} for an
excellent introduction to the geometry of K3 surfaces). By the Kummer construction, one obtains an
example of a K3 surface by means of a $\mathbb Z_2$-orbifold procedure from every complex $2$-torus.
On the other hand,  $\mathbb Z_2$-orbifolds of the 
toroidal CFTs are also reasonably well understood. 
One therefore expects to be able to construct examples
of CFTs which  allow a non-linear sigma model interpretation on some K3 surface. Compared
to CFTs on higher-dimensional Calabi-Yau $D$-manifolds, those on
K3 surfaces indeed provide a borderline case, in the sense
that much more is known about these so-called \textsc{K3 theories}. Most importantly,
we can give a mathematical definition of such theories without ever mentioning non-linear sigma
model constructions. The current section  presents this definition and summarizes some of
the known properties of K3 theories.\\

To motivate the mathematical definition of K3 theories, let us recall 
the conformal field theoretic elliptic genus of Section \ref{ellgen}.
Here we assume that we are given an $N=(2,2)$ superconformal field theory that obeys the following 
conditions, which are necessary for the CFT to allow a non-linear sigma model interpretation on
some 
Calabi-Yau $2$-manifold: the theory is superconformal at central charges $c=6,\,\overline c=6$ with space-time
supersymmetry, and such that all eigenvalues of $J_0$ and $\overline J_0$ are integral. 
This latter condition  is equivalent to the assumption that in addition to the spectral flow operator
of Ingredient \ref{stSUSY} in Section \ref{cftdef},
the theory possesses a quartet of 
\textsc{two-fold 
left- and right-handed spectral flow operators}
$\Theta^\pm,\,\overline\Theta^\pm$. By this we mean that these operators act
analogously to $\Theta^{\pm2}$ on the space of states,
with $\Theta$ as in (\ref{spfl}), namely
$$
\begin{array}{rclrcl}
[L_0,\Theta^\pm] &=&{c\over6}\Theta^\pm \mp\Theta^\pm\circ J_0,\quad&
[J_0,\Theta^\pm] &=&\mp{c\over3}\Theta^\pm ,\quad\\[5pt]
[\overline L_0,\overline \Theta^\pm] 
&=&{\overline c\over6}\overline \Theta^\pm \mp\overline \Theta^\pm\circ \overline J_0,\quad&
[\overline J_0,\overline \Theta^\pm] &=&\mp{\overline c\over3}\overline \Theta^\pm ,
\end{array}
$$
but with all other commutators vanishing.
The fields associated to $\Theta^\pm\Omega,\,\overline\Theta^\pm\Omega$ by the state-field correspondence
(Definition \ref{voa}) are denoted
$J^\pm(z)$ and $\overline J^\pm(\overline z)$, respectively.
By Proposition \ref{ellgenprop}, the conformal field theoretic elliptic genus $\mathcal E(\tau, z)$
of such a CFT is a weak Jacobi form of weight $0$ 
and index $1$. 
However, the space of such Jacobi forms is one-dimensional, as follows from the methods introduced in \cite{eiza85}
(see \cite{boli00} or \cite[Thm.~3.1.12]{diss} for direct proofs). According
to the discussion that follows Definition \ref{geoellgen}, 
the (complex) elliptic genus $\mathcal E_{\rm K3}(\tau, z)$ 
of a K3 surface is a weak Jacobi form of weight $0$ and index $1$ as well, which  by (\ref{ellgeninterpolates})
is non-zero, 
since $\mathcal E_{\rm K3}(\tau, z=0)=\chi({\rm K3})=24$. The precise form of the function 
$\mathcal E_{\rm K3}(\tau, z)$ is well-known, and we obtain
\begin{equation}\label{sameellipticgenera}
\mathcal E(\tau, z) 
= a\cdot \mathcal E_{\rm K3}(\tau, z) = a\cdot\left(2y+20+2y^{-1}+\mathcal O(q)\right)
\end{equation}
for some constant $a$. In fact, 
\begin{svgraybox}\vspace*{-1.5em}
\begin{proposition}[{\cite[\S7.1]{diss}}] \label{twoellgens}
Consider an $N=(2,2)$ superconformal field theory at central charges $c=6,\, \overline c=6$ with space-time supersymmetry
and such that all the eigenvalues of $J_0$ and of $\overline J_0$ are integral. 
\begin{enumerate}
\item
The elliptic genus of this CFT
either vanishes, or it agrees with the complex elliptic genus $\mathcal E_{\rm K3}(\tau,z)$ of a K3 surface.
\item
The conformal field theoretic elliptic genus vanishes if and only if the theory is a  toroidal $N=(2,2)$
superconformal field theory according to Definition \mbox{\rm\ref{susytoroidaldef}}.
\end{enumerate}
\end{proposition}
\vspace*{-1em}\end{svgraybox}
This result is mentioned in \cite{nawe00} and proved in \cite[\S7.1]{diss}, where the proof 
however contains a few typos. 
The sketch of a corrected proof is  banned to the Appendix, since it
uses a number of properties of superconformal field theories
with space-time supersymmetry which are well-known to the experts, but which we have not
derived in this exposition. \\

While as mentioned before, the toroidal $N=(2,2)$ superconformal field theories are well 
understood, it is also not hard to find examples of theories whose conformal field theoretic elliptic genus
is $\mathcal E_{\rm K3}(\tau, z)$, see  \cite{eoty89}. In particular, the authors 
of \cite{eoty89} prove  that the standard
$\mathbb Z_2$-orbifold of every toroidal $N=(2,2)$ superconformal field theory at central 
charges $c=6,\,\overline c=6$ yields such an example. 
By the above this is in accord with the expectations based on the Kummer construction,
hence our
\begin{svgraybox}\vspace*{-1.5em}
\begin{definition}\label{defk3cft}
A superconformal field theory is called a \textsc{K3 theory}, if the following
conditions hold: the CFT is an $N=(2,2)$ superconformal field theory at
 central charges $c=6,\, \overline c=6$ with space-time supersymmetry, 
all the eigenvalues of $J_0$ and of $\overline J_0$ are integral,
and the conformal field theoretic elliptic genus of the theory is
$$
\mathcal E(\tau, z)=\mathcal E_{\rm K3}(\tau, z).
$$
\end{definition}\vspace*{-1em}
\end{svgraybox}
Possibly, every K3 theory allows a non-linear sigma model interpretation
on some K3 surface, however a proof is far out of reach. Nevertheless, under 
standard assumptions on the deformation theory of such theories
it is possible to determine 
the form of every connected component of the moduli space of K3 theories. 
Namely, one assumes that all deformations by so-called marginal operators are integrable for
these theories, an assumption
which can be justified in string theory and which is demonstrated 
to all orders of perturbation theory in \cite{di87}. Then, based on the previous results \cite{se88,ce91},
one obtains
\begin{svgraybox}\vspace*{-1.5em}
\begin{theorem}[\cite{asmo94,nawe00}] \label{defk3cftmodspace}
With the notations introduced in Theorem \mbox{\rm\ref{toroidalmodulispace}},  
let $\mathcal T^{4,20}$ denote the Grass\-mannian of maximal positive definite 
oriented subspaces of $\mathbb R^{4,20}$,
$$
\mathcal T^{4,20}:= \mbox{O}^+(4,20;\mathbb R)\slash \mbox{SO}(4)\times \mbox{O}(20).
$$
By $\mathcal T^{4,20}_0\subset\mathcal T^{4,20}$ we denote the set of all those 
 maximal positive definite oriented subspaces $x\subset\mathbb R^{4,20}$ which have
the property that $x^\perp$ does not contain any \textsc{roots}, that is, all
$\alpha\in x^\perp\cap\mathbb Z^{4,20}$ obey $\langle\alpha,\alpha\rangle\neq-2$.

If the above-mentioned assumptions on  deformations of K3 theories  hold, namely that
all deformations by so-called marginal operators are integrable, then each
connected component $\mathcal M^{\rm K3}_s$
of the moduli space of K3 theories has the following form:
$$
\mathcal M^{\rm K3}_s=
\mbox{O}^+(4,20;\mathbb Z)\backslash \mathcal T^{4,20}_0.
$$
\end{theorem}
\vspace*{-1em}\end{svgraybox}
This result  reinforces the expectation that one connected component $\mathcal M^{\rm K3}_\sigma$
of the moduli space of K3 theories can be identified with the space of non-linear sigma models
on K3 surfaces, since in addition, we have
\begin{svgraybox}\vspace*{-1.5em}
\begin{proposition}[\cite{asmo94}]\label{geodef}
The partial completion $\mathcal T^{4,20}$ of the smooth universal covering space  
$\mathcal T_0^{4,20}$  of $\mathcal M^{\rm K3}_s$ can
be isometrically  identified with the \textsc{parameter space of non-linear sigma models on $K3$}.
Namely, denoting by $X$ the diffeomorphism type of a $K3$ surface, $\mathcal T^{4,20}$
is a cover of the space of triples $(\Sigma,\,V,\,B)$ where $\Sigma$ denotes a \textsc{hyperk\"ahler structure} on $X$,
$V\in\mathbb R^+$ is interpreted as the \textsc{volume} of $X$, 
and $B$ is the de Rham cohomology class of a real closed two-form on $X$,
a so-called \textsc{B-field}.

If a K3 theory in $\mathcal M^{\rm K3}_s$ lifts to a point in $\mathcal T^{4,20}$ which is mapped
to the triple $(\Sigma,\,V,\,B)$, then $(\Sigma,\,V,\,B)$ is called a
\textsc{geometric interpretation} of the K3 theory.
\end{proposition}
\vspace*{-1em}\end{svgraybox}
In \cite{nawe00,we00} it is shown that the
expectation that non-linear sigma models on K3 yield K3 theories indeed is
compatible with orbifold constructions, more precisely with
every orbifold construction of a K3 surface from a
complex two-torus by means of a discrete
subgroup of $\mbox{SU}(2)$. As mentioned above,
one might conversely expect that every  K3 theory with geometric interpretation $(\Sigma,\,V,\,B)$
can be constructed as a non-linear sigma model  on a K3 surface, specified
by the  data $(\Sigma,\,V,\,B)$. At least the 
existence of a non-linear sigma model interpretation has not been disproved
for any K3 theory, so far.\\

The statement of Proposition \ref{geodef} makes use of the fact that every K3 surface 
is a hyperk\"ahler manifold. The analogous statement for K3 theories is the observation
that the two commuting copies of $N=2$ superconformal  algebras (\ref{N=2a})--(\ref{N=2o}) 
are each extended to an \textsc{$N=4$ superconformal  algebra} in these theories. This is 
a direct consequence of our Definition 
\ref{defk3cft} of K3 theories. Indeed,  as mentioned at the beginning of this section,
the assumption of space-time supersymmetry together with the
integrality of the eigenvalues of $J_0$ and $\overline J_0$ 
imply that the fields  $J^\pm(z),\, \overline J^\pm(\overline z)$ corresponding to two-fold 
left- and right-handed spectral flow 
are fields of the CFT. One checks that at central charges
$c=6,\,\overline c=6$, these fields create states in the subspaces $W_1$ and $\overline W_1$ of
the vectorspaces underlying the chiral algebras of Property \ref{chiral}
(see the vacuum axiom in Definition \ref{voa}), whose $J_0$- (respectively $\overline J_0$-) eigenvalues
are $\pm2$. Moreover, with the $U(1)$-currents $J(z),\, \overline J(\overline z)$
of the two commuting copies of 
$N=2$ superconformal vertex algebras, the fields $J^\pm(z),\, \overline J^\pm(\overline z)$
generate two commuting copies of a so-called \textsc{$\mathfrak{s}\mathfrak{u}(2)_1$-current algebra},
which in turn is known to extend the $N=2$ superconformal  algebra to an $N=4$ superconformal
algebra \cite{aetal76}.

The characters of the irreducible unitary representations of the 
relevant $N=4$ superconformal algebra at arbitrary
central charges have been determined in \cite{egta87,egta88a,egta88,egta88b,ta89}. Their 
transformation properties under modular transforms in general are \textsl{not} modular, 
in contrast to the situation at lower supersymmetry, where 
an infinite class of
characters of irreducible unitary representations 
does enjoy modularity. Instead, these $N=4$ characters 
exhibit a so-called \textsc{Mock modular} behavior,
see e.g. \cite{dmz12} for a recent account. Since in the context of non-linear sigma models, 
$N=4$ supersymmetry is linked to the
geometric concept of hyperk\"ahler manifolds \cite{agf81}, this seems to point towards a connection between Mock 
modularity and hyperk\"ahler geometry. The nature of this connection however, to date, is completely 
mysterious.
\section{The elliptic genus of K3}\label{ellk3}
Recall that
the elliptic genus $\mathcal E_{\rm K3}(\tau, z)$
of K3 plays center stage in our Definition \ref{defk3cft} of K3 theories. Though this function 
is explicitly known and well understood, recent years have uncovered a number of mysteries
around it. In the present section, some of these mysteries are discussed. This involves 
more open than solved problems, and as a reminder, the titles of all the following subsections are 
questions instead of statements.
\subsection{A non-geometric decomposition of the elliptic genus?}\label{ellk3deco}
As was mentioned at the end of Section \ref{k3cft}, our very Definition \ref{defk3cft} ensures
that every K3 theory enjoys $N=(4,4)$ supersymmetry. The current section summarizes
how this induces a decomposition of the function $\mathcal E_{\rm K3}(\tau, z)$, which is a priori not
motivated geometrically and which turns out to bear some intriguing surprises.\\

In what follows, assume that we are given 
a K3 theory according to Definition \ref{defk3cft} with space of states 
$\mathbb H=\mathbb H^{NS}\oplus\mathbb H^R$. Both $\mathbb H^{NS}$ and $\mathbb H^R$
can be decomposed into direct sums of irreducible unitary
representations with respect to the $N=(4,4)$ superconformal symmetry.
According to \cite{egta87,egta88a}, there are three types of irreducible
unitary representations of the relevant $N=4$ superconformal algebra at central charge $c=6$, namely
the \textsc{vacuum representation},
the \textsc{massless matter representation},
and finally the \textsc{massive matter representations} which form a one-parameter
family indexed by $h\in\mathbb R_{>0}$. For later convenience we 
focus on the Ramond-sector $\mathbb H^R$ of  our theory and 
denote the respective irreducible unitary representations by
$\mathcal H_0$, $\mathcal H_{\rm mm}$, $\mathcal H_h$ ($h\in\mathbb R_{>0}$).
This notation alludes to
the properties of the corresponding representations
in the Neveu-Schwarz sector $\mathbb H^{NS}$, which are related to the representations
in $\mathbb H^R$ by spectral flow $\Theta$ according to (\ref{spfl}). Indeed, the vacuum representation in the NS-sector 
has the vacuum $\Omega$ as its ground state. The massive matter representations are
characterized by the spontaneous breaking of  supersymmetry at every mass level,
including the ground state \cite{wi82}.

Setting $y=\exp(2\pi i z)$ and $q=\exp(2\pi i\tau)$ for $z,\,\tau\in\mathbb C$ with $\Im(\tau)>0$
as before and using $c/24=1/4$, the  characters of the irreducible unitary $N=4$
representations that are relevant to our discussion
are denoted  by
$$
\chi_a (\tau,z) := \mbox{Str}_{\mathcal H_a} \left(  y^{J_0} q^{L_0-1/4} \right)
= \mbox{Tr}_{\mathcal H_a} \left( (-1)^{J_0} y^{J_0} q^{L_0-1/4} \right),\;\;
a\in\mathbb R_{\geq0}\cup\{{\rm mm}\}.
$$
These functions have been determined explicitly in \cite{egta88a}.
For our purposes, only the following properties are relevant, %
\begin{equation}\label{wittenindices}
\begin{array}{rclrcl}
\chi_0(\tau, z=0) &=& -2,  &\chi_{\rm mm}(\tau, z=0) &=& 1, \quad \\[5pt]
\forall h>0\colon\;\;\chi_h(\tau,z) &=& q^h \widehat\chi(\tau,z)&
\mbox{ with }\quad
\widehat\chi(\tau,z) &=&\chi_0(\tau, z) + 2\chi_{\rm mm}(\tau, z),\\[5pt]
&&&\mbox{hence }\quad
\chi_h(\tau, z=0) &=&  \widehat\chi(\tau, z=0)\;=\;0.
\end{array} 
\end{equation}
The constant $\chi_a (\tau,z=0)$ yields  the so-called \textsc{Witten index}  \cite{wi82,wi87,wi88}
of the respective representation.

The most general ansatz for a decomposition of $\mathbb H^R$ into irreducible representations of
the two commuting $N=4$ superconformal algebras therefore reads
$$
\mathbb H^R = \bigoplus_{a,\,\overline a\in\mathbb R_{\geq0}\cup\{{\rm mm}\}} 
m_{a,\overline a} \mathcal H_a\otimes \overline{\mathcal H_{\overline a}}
$$
with appropriate non-negative integers $m_{a,\overline a}$. Then
$$
\mbox{Tr}_{\mathbb H^R} \left( (-1)^{J_0-\overline J_0} y^{J_0} \overline y^{\overline J_0}
q^{L_0-1/4} \overline q^{\overline L_0-1/4} \right) 
= \sum_{a,\,\overline a\in\mathbb R_{\geq0}\cup\{{\rm mm}\}} 
m_{a,\overline a}\cdot \chi_a(\tau,z)\cdot \overline{\chi_{\overline a}(\tau, z)},
$$
together with Definition \ref{cftellgen} yields the conformal field theoretic elliptic genus of our CFT as
\begin{equation}\label{ellgenform}
\mathcal E(\tau,z) = \sum_{a,\,\overline a\in\mathbb R_{\geq0}\cup\{{\rm mm}\}} 
m_{a,\overline a}\cdot \chi_a(\tau,z)\cdot \overline{\chi_{\overline a}(\tau, z=0)}.
\end{equation}
This expression  simplifies dramatically on insertion of (\ref{wittenindices}). In addition,
the known properties of K3 theories impose a number of constraints on the coefficients 
$m_{a,\overline a}$. First, since under spectral flow,  $\mathcal H_0$ is mapped
to the irreducible representation of the $N=4$ superconformal algebra whose ground state is
the vacuum $\Omega$, the uniqueness of the vacuum (see  Property \ref{chiral}) 
implies $m_{0,0}=1$. Moreover,  from the proof of Proposition \ref{twoellgens} (see the Appendix) 
or from the known explicit form (\ref{sameellipticgenera}) of $\mathcal E(\tau,z)$,
we deduce that in every K3 theory,  $m_{0,{\rm mm}}=m_{{\rm mm},0}=0$. 
Finally, according to the discussion of Property \ref{partfun} in Section \ref{cftdef}, 
$\mathbb H_{h,\overline h}\cap\mathbb H_b$ can only be non-trivial if $h-\overline h\in\mathbb Z$,
which on $\mathbb H_{h,\overline h}\cap\mathbb H_f\cap\mathbb H^{NS}$ 
generalizes to $h-\overline h\in{1\over2}+\mathbb Z$.
Since the groundstates of $\mathcal H_0,\,\mathcal H_{\rm mm},\, \mathcal H_h$ under spectral flow
yield  states with $L_0$-eigenvalues $0,\,{1\over2},\, h$, 
and $J_0$-eigenvalues $0,\,\pm1,\,0$, respectively \cite{egta88a}, this implies that
$m_{0,\overline h},\,m_{{\rm mm},\overline h},\,m_{h,0},\,m_{h,{\rm mm}}$ with $h,\,\overline h>0$
can only be non-zero if $h,\,\overline h\in\mathbb N$. In conclusion, we  obtain a refined ansatz for the $N=(4,4)$
decomposition of $\mathbb H^R$,
\begin{equation}\label{refinedansatz}
\begin{array}{rcl}
\mathbb H^R = \mathcal H_0\otimes \overline{\mathcal H_0}
&\oplus& \displaystyle
h^{1,1} \mathcal H_{\rm mm}\otimes \overline{\mathcal H_{\rm mm}}
\oplus \bigoplus_{h,\, \overline h\in\mathbb R_{>0} } k_{h,\overline h} \mathcal H_h\otimes \overline{\mathcal H_{\overline h}}\\[5pt]
&&\oplus\displaystyle \bigoplus_{n=1}^\infty 
\left[f_n \mathcal H_n\otimes \overline{\mathcal H_0}\oplus \overline{f_n} \mathcal H_0\otimes \overline{\mathcal H_n}\right]\\[5pt]
&\oplus&\displaystyle \bigoplus_{n=1}^\infty  
\left[g_n \mathcal H_n\otimes \overline{\mathcal H_{\rm mm}}\oplus \overline{g_n} \mathcal H_{\rm mm}\otimes \overline{\mathcal H_n}\right].
\end{array}
\end{equation}
Here, all the coefficients
$h^{1,1},\, k_{h,\overline h},\, f_n,\,\overline f_n,\, g_n,\,\overline g_n$ are non-negative integers,
whose precise values depend  on the specific K3 theory under inspection.
By  (\ref{ellgenform}), and inserting (\ref{wittenindices}) and the
refined ansatz (\ref{refinedansatz}), we  obtain
\begin{eqnarray*}
\mathcal E(\tau,z) 
&=&-2 \chi_0(\tau,z) +  h^{1,1} \chi_{\rm mm}(\tau,z) + \sum_{n=1}^\infty \left[-2f_n +g_n\right]\chi_n(\tau,z) \\
&=&-2 \chi_0(\tau,z) +  h^{1,1} \chi_{\rm mm}(\tau,z) + e(\tau)\,\widehat\chi(\tau,z)
\mbox{ with }
e(\tau):= \sum_{n=1}^\infty \left[g_n-2f_n\right]q^n.
\end{eqnarray*}
Now recall from Definition \ref{defk3cft} that $\mathcal E(\tau,z) =\mathcal E_{\rm K3}(\tau,z)$, where by the
discussion preceding (\ref{sameellipticgenera}) we have $\mathcal E_{\rm K3}(\tau,z=0)=24$.
Using (\ref{wittenindices}),
this implies $h^{1,1}=20$. Since the complex elliptic genus 
$\mathcal E_{\rm K3}(\tau,z)$ is a topological invariant of all K3 surfaces, 
we conclude
\begin{svgraybox}\vspace*{-1.5em}
\begin{proposition}\label{definitionofe}
The elliptic genus $\mathcal E_{\rm K3}(\tau,z)$ of K3 decomposes into the characters of
irreducible unitary representations of the 
relevant $N=4$ superconformal algebra in the Ramond sector 
according to
\begin{eqnarray*}
\mathcal E_{\rm K3}(\tau,z) 
&=&-2 \chi_0(\tau,z) +  20 \chi_{\rm mm}(\tau,z) + e(\tau)\,\widehat\chi(\tau,z),\\
&&\quad\quad\quad\quad\quad\mbox{ where }\;
e(\tau):= \sum_{n=1}^\infty \left[g_n-2f_n\right]q^n,
\end{eqnarray*}
and the coefficients $g_n$, $f_n$ give the respective multiplicities of representations in the
decomposition (\ref{refinedansatz}). While the values of $g_n$, $f_n$  vary 
within the moduli space of K3 theories, the coefficients $g_n-2f_n$ of $e(\tau)$ are invariant. 
\end{proposition}
\vspace*{-1em}\end{svgraybox}
A decomposition of $\mathcal E_{\rm K3}(\tau,z)$
in the spirit of Proposition \ref{definitionofe} was already given in \cite{eoty89}. 
In \cite{oo89} and independently in
\cite[Conj.7.2.2]{diss}  it was  conjectured that all coefficients of the function $e(\tau)$ are non-negative,
for the following reason:
recall that under spectral flow, the irreducible representation $\mathcal H_0$ is mapped
to the representation of the $N=4$ superconformal algebra whose ground state is
the vacuum $\Omega$. Therefore,  
in (\ref{refinedansatz}), the coefficients $f_n$ determine those contributions to
the subspace $W_n\subset W$ of the vectorspace underlying
the chiral algebra of Property \ref{chiral} that
do \textsl{not} belong to the vacuum representation under the  $N=(4,4)$
supersymmetry. For any fixed value of $n\in\mathbb N$ with $n>0$, we 
generically expect no such additional contributions
to $W_n$. In other words, we expect that generically $f_n=0$ and thus that the $n^{\rm th}$ coefficient of $e(\tau)$ agrees with
$g_n\geq0$. Since these coefficients are invariant on the moduli space of K3 theories, they should
always be non-negative.

The conjectured positivity of the coefficients $g_n-2f_n$ is proved in \cite{eghi09,eot10}
in the context of an intriguing observation. Namely, in \cite{eot10},  Eguchi, Ooguri and Tachikawa  
observe that each of these coefficients  seems to give the
dimension of a representation of a certain sporadic group, namely of the \textsc{Mathieu group
$M_{24}$}. 
For small values of $n$, they find dimensions of irreducible representations, while at higher order, 
more work is required to arrive at a well-defined conjecture.
The quest for understanding this observation,
which is often referred to as \textsc{Mathieu Moonshine}, has sparked enormous interest in the 
mathematical physics community.
Building on results of \cite{ch10,ghv10b,ghv10a,eghi11}, the observation has been
recently verified by Gannon
in the following form:
\begin{svgraybox}\vspace*{-1.5em}
\begin{theorem}[\cite{ga12}]\label{gannonresult}
There are  virtual representations of the Mathieu group $M_{24}$ on spaces
$\mathcal R_0$, $\mathcal R_{\rm mm}$,
and true representations  on spaces $\mathcal R_n$,
$n\in\mathbb N_{>0}$, such that 
$$
\mathcal R := \mathcal H_0\otimes \mathcal R_0
\oplus  \mathcal H_{\rm mm}\otimes \mathcal R_{\rm mm}
\oplus  \bigoplus_{n=1}^\infty \mathcal H_n\otimes \mathcal R_n
$$
has the following properties:
with the $N=4$ superconformal algebra acting non-trivially only 
on the first factor in each summand of $\mathcal R$, 
and the Mathieu group $M_{24}$ acting non-trivially only  on the second factor,
one obtains functions 
$$
\forall g\in M_{24}\colon\quad
\mathcal E_g(\tau,z):= \mbox{Tr}_{\mathcal R} \left( (-1)^{J_0}g y^{J_0} q^{L_0-1/4}\right)
$$
which
under modular transformations
generate a collection of \textsc{$M_{24}$-twisted elliptic genera} of K3. 
In particular, 
$\mathcal E_{\rm id}(\tau,z)=\mathcal E_{\rm K3}(\tau,z)$.
\end{theorem}
\vspace*{-1em}\end{svgraybox}
\subsection{A geometric Mathieu Moonshine phenomenon?}\label{symmk3}
While Theorem \ref{gannonresult} beautifully specifies a well-defined formulation of the 
Mathieu Moonshine observation and 
proves it, the proof does not offer any insight into
the role of the Mathieu group $M_{24}$ in the context of K3 theories.
The present section summarizes some ideas for a possible interpretation
that is based in geometry.\\

Indeed, the relevance of the group $M_{24}$ for the geometry of K3 surfaces
had  been discovered much earlier by Mukai:
\begin{svgraybox}\vspace*{-1.5em}
\begin{theorem}[\cite{mu88}]\label{mukaitheorem}
Let $G$ denote a finite group of \textsc{symplectic automorphisms} of a K3 surface $X$.
By this we mean that $X$ denotes a K3 surface whose complex structure has been fixed,
and that $G$ is a finite group of biholomorphic maps on $X$ whose induced action on the 
holomorphic volume form is trivial.

Then $G$ is a subgroup of the Mathieu group $M_{24}$.
More precisely, $G$ is a subgroup of one out of a list of $11$ subgroups of
$M_{23}\subset M_{24}$, the largest of which has order $960$.
\end{theorem}
\vspace*{-1em}\end{svgraybox}
Hence although
$M_{24}$ does play a crucial role in describing symplectic automorphisms of K3 surfaces,
Theorem \ref{mukaitheorem} cannot immediately explain Mathieu Moonshine.
Indeed,  Mathieu Moonshine
suggests that there is an action of the entire group $M_{24}$ on some mathematical object which underlies
the elliptic genus of K3, while Theorem \ref{mukaitheorem} implies that no
K3 surface allows $M_{24}$ as its symplectic automorphism group. Namely,
 the theorem states
 that the maximal order of a 
symplectic automorphism group of any K3 surface
is $960$, which is smaller by orders of magnitude than
the order $244.823.040$ of $M_{24}$.\\

Since the \textsl{non-geometric} decomposition of $\mathcal E_{\rm K3}(\tau,z)$ 
by means of $N=4$ supersymmetry presented in Section \ref{ellk3deco}
led to the discovery of Mathieu Moonshine, one may suspect 
that rather than the properties of K3 surfaces, the properties of K3 theories should 
explain the Mathieu Moonshine phenomena.
However, symmetry groups of K3 theories, in general, need not be subgroups of $M_{24}$,
as apparently was  first noted independently by the authors of \cite{eot10} and \cite{tawe11}.
Conversely, no K3 theory can
have $M_{24}$ as its symmetry group, as follows from  \cite{ghv12},
where Gaberdiel, Hohenegger and Volpato 
generalize a very enlightening second proof of Theorem \ref{mukaitheorem} due to  
Kondo \cite{ko98} to a classification result for symmetries of K3 theories.\\

Because by the above, the symmetries of K3 theories seem not to explain Mathieu Moonshine,
in a series of papers  \cite{tawe11,tawe12,tawe13}, 
Taormina and the author have proposed constructions 
that may lead to
the action of $M_{24}$  as a \textsl{combined action} of all finite symplectic
automorphism groups of K3 surfaces.
This idea can be motivated by the mathematical properties of the elliptic genus
which were presented in Sections \ref{ellgen} and \ref{chiderham}:

By Theorem \ref{voafromchdr}, the complex elliptic genus $\mathcal E_{\rm K3}(\tau,z)$
is recovered from the chiral de Rham complex $\Omega_X^{\rm ch}$
of a K3 surface $X$ as its graded Euler characteristic,
that is, 
as the supertrace of the appropriate operator on the sheaf  cohomology
$H^\ast(X,\Omega_X^{\rm ch})$.
In accord with \cite[(2.1.3)]{goma03}, one can expect that
every symplectic automorphism of a K3 surface $X$ induces an action on 
 $H^\ast(X,\Omega_X^{\rm ch})$.
 Therefore, the sheaf cohomology $H^\ast({\rm K3},\Omega_{\rm K3}^{\rm ch})$ of the chiral de Rham
complex appears to be an excellent candidate for 
the desired mathematical object which both underlies the elliptic genus,
and which carries actions of all finite symplectic automorphism groups of K3 surfaces,
thus combining them to the action of a possibly larger group.

Note that according to Theorem \ref{voafromchdr}, there exists a natural structure of a supervertex algebra on
$H^\ast({\rm K3},\Omega_{\rm K3}^{\rm ch})$. This additional structure on the 
mathematical object which underlies the elliptic genus is in complete
accord with the implications of Theorem \ref{gannonresult}. Indeed, it was already conjectured
in \cite{gprv12,gpv13}, that Mathieu Moonshine is governed by some vertex algebra
which carries an $M_{24}$-action, whose properties would immediately induce the
modular transformation properties of the twisted elliptic genera of Theorem \ref{gannonresult}.
As was argued in the discussion of Theorem \ref{voafromchdr},  $H^\ast(X,\Omega_X^{\rm ch})$
is moreover expected to be related to  a non-linear sigma model on $X$, at least in an infinite
volume limit, providing the desired link to K3 theories. Indeed, the required compatibility with an
infinite volume limit might also explain the restriction to those symmetries of K3 theories which 
can be induced by some symplectic automorphism 
of finite order of a K3 surface, as seems to be the case for the 
generators of $M_{24}$ in Mathieu Moonshine.

Unfortunately, despite all its convincing properties promoting it to an excellent candidate
to resolve Mathieu Moonshine, the vertex algebra structure of 
$H^\ast(X,\Omega_X^{\rm ch})$ is notoriously hard to calculate, 
as are the precise properties of general non-linear sigma models on K3, even in an infinite volume limit.
Therefore, these ideas remain conjectural, so far.
Sadly, known alternative  constructions for vertex algebras that underlie the elliptic genus
and that are easier to calculate
seem not to explain Mathieu Moonshine  \cite{crho13}.\\

The above-mentioned mechanism of combining symplectic automorphism groups of distinct
K3 surfaces to larger groups, as presented in \cite{tawe11,tawe12}, yields the
following result, which can be seen as evidence in favor of these ideas:
\begin{svgraybox}\vspace*{-1.5em}
\begin{proposition}[\cite{tawe13}]\label{taorminawendlandresult} 
Consider the ``smallest massive''
representation of $M_{24}$ that occurs in Theorem \mbox{\rm\ref{gannonresult}}, 
that is, the representation on $\mathcal R_1$.

The space $\mathcal R_1$ is isomorphic to a  certain vectorspace $V^{CFT}$ of states which is common to
all K3 theories that are obtained by a standard $\mathbb Z_2$-orbifold construction
from a toroidal $N=(2,2)$ superconformal field theory. Moreover, 
on $V^{CFT}$, the combined action of all 
finite symplectic automorphism groups of Kummer surfaces induces
a faithful action of the maximal subgroup $\mathbb Z_2^4\rtimes A_8$ of order 
$322.560$ in $M_{24}$.
The resulting representation on $V^{CFT}$
is equivalent to the representation of $\mathbb Z_2^4\rtimes A_8$ on $\mathcal R_1$
which is induced by restriction from $M_{24}$ to this subgroup.
\end{proposition}
\vspace*{-1em}
\end{svgraybox}
This result is the first piece of evidence in the literature for any trace of an
$M_{24}$-action on a space of states of a K3 theory. It is remarkable that the
CFT techniques produce precisely the representation of a maximal subgroup of
$M_{24}$ which is predicted by Mathieu Moonshine according to the idea of
``combining symplectic automorphism groups''.
Note that the group $\mathbb Z_2^4\rtimes A_8$ is not a subgroup of $M_{23}$, 
indicating that indeed $M_{24}$ rather than $M_{23}$ should be expected to be responsible for
Mathieu Moonshine, despite Theorem \ref{mukaitheorem}, by which all finite
symplectic automorphism groups of K3 surfaces are subgroups of $M_{23}$.
This preference for $M_{24}$ to $M_{23}$
is in accord with the findings of \cite{ga12}.

Encouraged by Proposition \ref{taorminawendlandresult},
one may hope that in an infinite volume limit, $V^{CFT}$
can be identified with a subspace of 
$H^\ast({\rm K3},\Omega_{\rm K3}^{\rm ch})$,
inducing an equivalence of vertex algebras.
Furthermore, by combining the action of $\mathbb Z_2^4\rtimes A_8$ with the action of
finite symplectic automorphism groups of K3 surfaces which are not Kummer,
one may hope to generate an action of the entire group $M_{24}$.
Finally, one may hope that
this result generalizes to the remaining representations on $\mathcal R_n$, $n>1$,
found in Theorem \ref{gannonresult}. 
In conclusion, there is certainly much work left.
\subsection{A geometric decomposition of the elliptic genus?}\label{myconjecture}
Even if the ideas presented in Section \ref{symmk3} prove successful, then
so far, they give no indication for the reason for $M_{24}$ -of all groups- to arise from
 the combined action of finite symplectic automorphism groups
of K3 surfaces. 
Circumventing this intrinsic problem,
the current section presents a simpler conjecture which can be formulated independently of
Mathieu Moonshine. If true, however, it could serve as a step towards understanding 
Mathieu Moonshine.\\

Taking the idea seriously that there should be a purely geometric explanation for
Mathieu Moonshine, one main obstacle to unraveling its mysteries 
is the lack of geometric interpretation for the non-geometric decomposition of
$\mathcal E_{\rm K3}(\tau,z)$ stated in Proposition \ref{definitionofe}, which is at the heart of the
discovery of Mathieu Moonshine. Recall that the derivation of Proposition \ref{definitionofe} rests on the identification
(\ref{mckeantype}) of the complex elliptic genus of a 
 Calabi-Yau $D$-manifold $X$
with the conformal field theoretic elliptic genus of a CFT that is obtained 
from $X$ by a non-linear sigma model construction.  
We have incorporated this identification into our Definition \ref{defk3cft}, and it is
the  motivation for
decomposing the complex elliptic genus of K3 into the
characters of irreducible unitary representations of the 
relevant $N=4$ superconformal algebra at
central charge $c=6$.
While the conformal field theoretic elliptic genus by Definition \ref{cftellgen}
is obtained as a trace over the
space of states $\mathbb H^R$, the complex elliptic genus by Definition \ref{geoellgen}
is an analytic trace over a formal power series $\mathbb E_{q,-y}$ whose coefficients are 
holomorphic vector bundles on our K3 surface.
The decomposition of the space of states $\mathbb H^R$ of every K3 theory by 
$N=(4,4)$ supersymmetry which was performed in Section \ref{ellk3deco} 
to derive Proposition \ref{definitionofe} should 
accordingly be counterfeited by a decomposition of  $\mathbb E_{q,-y}$.
We thus expect
\begin{svgraybox}\vspace*{-1.5em}
\begin{conjecture}\label{localconjecture}
Let $X$ denote a K3 surface with holomorphic tangent bundle $T:=T^{1,0}X$, and consider 
$\mathbb E_{q,-y}$ as in Definition \ref{geoellgen}. Furthermore, let 
$e(\tau)$ denote the function 
defined in Proposition \ref{definitionofe}.
Then there are polynomials $p_n$, $n\in\mathbb N_{>0}$, such that
\begin{eqnarray*}
\mathbb E_{q,-y} 
&=& -\mathcal O_X\cdot \chi_0(\tau,z) -  T\cdot \chi_{\rm mm}(\tau,z) + \sum_{n=1}^\infty p_n(T)\cdot q^n\widehat\chi(\tau,z),\\
&&\mbox{and }\quad
e(\tau) = \sum_{n=1}^\infty \left(\int_X\mbox{Td}(X)p_n(T) )\right) \cdot q^n,
\end{eqnarray*}
where $p_n(T)=\smash{\bigoplus\limits_{k=0}^{N_n}} \alpha_k T^{\otimes k}$ if 
$p_n(x) = \smash{\sum\limits_{k=0}^{N_n}} \alpha_k x^k$, $\alpha_k\in\mathbb Z$, and where
$T^{\otimes 0}= \mathcal O_X$ is understood.
\end{conjecture}
\vspace*{-0.5em}
\end{svgraybox}
If (\ref{mckeantype}) is interpreted as a generalization of the McKean-Singer Formula,
as indicated in the discussion of that equation, then Conjecture \ref{localconjecture}
can be viewed as a generalization of a \textsc{local index theorem}
\cite{pa71,gi73,abp73,ge83}.
Note that the conjecture is formulated without even alluding to Mathieu Moonshine,
so it may be of independent interest.
If true, then  for each $n\in\mathbb N_{>0}$,
every finite symplectic automorphism group of a K3 surface
$X$ naturally acts on  $p_n(T)$, and one may hope that
this will yield insight into the descent of  this action  to the representation 
of $M_{24}$ on $\mathcal R_n$ which was found in Theorem \ref{gannonresult}.
\begin{acknowledgement}\hspace*{\fill}\\
It is my pleasure to thank Ron Donagi for his hospitality and 
for many inspiring discussions on elliptic genera and  Mathieu Moonshine
which have greatly influenced this work. My sincere thanks also go to Scott Carnahan, 
since the final touches to this exposition have benefited
much from my discussions with him. 

Some of the material underlying this exposition arose from lecture courses 
that I presented at two
summer schools, and I am grateful for the invitations to these events:
cordial thanks go to the
organizers Alexander Cardona, Sylvie Paycha, Andr\'es Reyes, Hern\'an Ocampo
and the participants
of the 8th Summer School on ``Geometric, Algebraic and Topological Methods
for Quantum Field Theory 2013'' at
Villa de Leyva, Columbia, 
as well as the organizers Martin Schlichenmaier, Pierre Bielavsky, Harald Grosse,
Ryoichi Kobayashi, Armen Sergeev, Oleg Sheinman, Weiping Zhang 
and the participants of the
ESI School and Conference ``Geometry and Quantization
GEOQUANT 2013'' at Vienna, Austria,
for creating such inspiring  events.

This work is in part  supported by my 
ERC Starting Independent Researcher Grant StG No. 204757-TQFT
``The geometry of topological quantum field theories''.
\end{acknowledgement}
\section*{Appendix: proof of Proposition \ref{twoellgens} in Section \ref{k3cft}}\label{twoellgensproof}
\addcontentsline{toc}{section}{Appendix}
The entire proof of Proposition \ref{twoellgens}
rests on the study of the $+1$-eigenspace of 
the linear operator $\overline J_0$ on
the subspace $\overline W_{1/2}$ of the vectorspace $\overline W$ underlying the chiral algebra. 
 First, one shows that this eigenspace
is either trivial or two-dimensional, and from this one deduces claim 1. of the proposition.
One direction of   claim 2. is checked by direct calculation, using the defining properties of toroidal
$N=(2,2)$ superconformal field theories.  To obtain the converse, one shows that 
$\mathcal E(\tau, z)\equiv0$ implies that an antiholomorphic counterpart of 
the conformal field theoretic elliptic genus must vanish as well, from which  claim 2. is shown to follow. 
\begin{enumerate}
\item
Assume that the space $\overline W_{1/2}$ contains an eigenvector of
$\overline J_0$ with eigenvalue $+1$. 
We denote the  field associated to this state 
by $\overline\psi^+_{1}(\overline z)$. 
The properties of the real structure on the space of states $\mathbb H$ of our CFT imply that there is a
complex conjugate state with $\overline J_0$-eigenvalue $-1$ whose 
associated field we denote by $\overline\psi^-_{1}(\overline z)$. The properties of unitary irreducible
representations of the Virasoro algebra imply that these fields form a  \textsc{Dirac fermion} (see
the discussion around (\ref{Dirac})), and  that therefore
$\overline J_1(\overline z):={1\over2}:\!\!\overline\psi^+_{1}\overline\psi^-_{1}\!\!\!:\!\!(\overline z)$ is a 
$U(1)$-current as in Example \ref{U1current}
in Section \ref{vertexalgebras}. By a procedure known as \textsc{GKO-construction} \cite{gko85}, one obtains 
$\overline J(\overline z)=\overline J_1(\overline z)+\overline J_2(\overline z)$ 
for the field $\overline J(\overline z)$ in the $N=2$ superconformal algebra
(\ref{N=2a})--(\ref{N=2o}), and $\overline J_k(\overline z)=i\partial\overline H_k(\overline z)$ with 
$\overline\psi^\pm_{1}(\overline z)
=c_1^\pm:\!e^{\pm i\overline H_1}\!\!:\!\!(\overline z)$
with cocycle factors $c_1^\pm$. The fields of twofold right-handed spectral flow,
which by assumption are fields of the theory, are moreover
given by  $\overline J^\pm(\overline z)
=c^\pm:\!e^{\pm i(\overline H_1+\overline H_2)}\!\!:\!\!(\overline z)$
with cocycle factors $c^\pm$. Their OPEs with the 
$\overline\psi^\pm_{1}(\overline z)$ yield an 
additional Dirac-fermion, with fields  $\overline\psi^\pm_{2}(\overline z):=
c_2^\pm:\!e^{\pm i\overline H_2}\!\!:\!\!(\overline z)$  in the CFT and further 
cocycle factors $c_2^\pm$.
This proves that the $\pm1$-eigenspaces of 
$\overline J_0$ on $\overline W_{1/2}$ each are precisely two-dimensional, since by the same argument no further 
Dirac fermions can be fields of the theory. Note that by definition, the corresponding states belong to the sector
$\mathbb H_f\cap\mathbb H^{NS}\subset\mathbb H$ of the space of states of our theory.

In summary, the $+1$-eigenspace of 
the linear operator $\overline J_0$ on
$\overline W_{1/2}$ is either trivial or two-dimensional.
 
We now  study the leading order contributions 
in the conformal field theoretic elliptic genus $\mathcal E(\tau,z)$ of our theory.
From (\ref{sameellipticgenera}) and by the very Definition \ref{cftellgen}
we deduce that $2ay^{-1}$ counts 
states in the subspace $V\subset\mathbb H^R$ where 
$L_0,\, \overline L_0$ both take eigenvalue  ${c\over24}={1\over 4}={\overline c\over24}$
and $J_0$ takes eigenvalue  $-1$. 
More precisely, 
$$
2a=\mbox{Tr}_V\left( (-1)^{J_0-\overline J_0} \right)=-\mbox{Tr}_V\left( (-1)^{\overline J_0} \right).
$$ 
As follows from properties of the so-called \textsc{chiral ring}, see e.g.\ \cite[\S3.1.1]{diss}, 
a basis of $V$ is obtained by spectral flow $\Theta$ (see Ingredient \ref{stSUSY} in Section \ref{cftdef})
from (i) the vacuum $\Omega$, (ii) the state whose corresponding field is $\overline J^+(\overline z)$, and (iii) 
a basis  of the $+1$-eigenspace of  the linear operator $\overline J_0$ on $\overline W_{1/2}$. 
Since according to (\ref{spfl}),
the eigenvalues of $\overline J_0$ after spectral flow to $V$ are (i) $-1$, (ii) $+1$, (iii) $0$,
the above trace vanishes if the $+1$-eigenspace of  the linear operator $\overline J_0$ on $\overline W_{1/2}$
is two-dimensional, implying $2a=0$,
and if this eigenspace is trivial, then we obtain $2a=2$. 

In conclusion, the conformal field theoretic elliptic genus of our theory either vanishes, in
which case the $+1$-eigenspace of  the linear operator $\overline J_0$ on $\overline W_{1/2}$
is created by two Dirac fermions, or $\mathcal E(\tau, z) 
=  \mathcal E_{\rm K3}(\tau, z)$.
\hspace*{\fill}\qed
\item
\begin{enumerate}
\item
Using the details of toroidal $N=(2,2)$ superconformal field theories that are summarized in Section \ref{toroidalcfts},
one checks by a direct calculation that the conformal field theoretic elliptic genus of all such theories vanishes.
\hspace*{\fill}\qed
\item
To show the converse, first observe that
in our discussion of $N=(2,2)$ superconformal field theories, the two commuting copies of a 
superconformal algebra are mostly treated on an equal level. However, the Definition \ref{cftellgen}
breaks this symmetry, and
$$
\overline{\mathcal E}(\tau, z) 
:= \mbox{Tr}_{\mathbb H^{R}}
\left( (-1)^{J_0-\overline J_0}\overline y^{\overline J_0} q^{L_0-c/24} \overline q^{\overline L_0-\overline c/24}\right)
$$
should define an equally important antiholomorphic counterpart of the conformal field theoretic elliptic genus.
In our case, by the same reasoning as for $\mathcal E(\tau,z)$, it must yield zero or 
$\overline{\mathcal E_{\rm K3}(\tau, z)}$.
Note that Proposition \ref{ellgenprop} implies that
$$
{\mathcal E}(\tau, z=0) =
\overline{\mathcal E}(\tau, z=0)
$$
is a constant, which in fact is known as the \textsc{Witten index} \cite{wi82,wi87,wi88}. 
In particular,  by (\ref{ellgeninterpolates}) we have $\mathcal E_{\rm K3}(\tau,z=0)=24$,  hence
$\mathcal E(\tau, z)\equiv0$ implies $\overline{\mathcal E}(\tau, z)\equiv0$.
It remains to be shown that our theory is a toroidal theory according to Definition \ref{susytoroidaldef}
in this case. 

But Step 1. of our proof then implies  that the $+1$-eigenspace of  the linear operator $\overline J_0$ on $\overline W_{1/2}$
is created by two Dirac fermions and that the analogous statement holds for the $+1$-eigenspace of  the linear operator 
$J_0$ on $W_{1/2}$. Hence we have Dirac fermions
$\psi_k^\pm(z)$ and $\overline\psi_k^\pm(\overline z)$, $k\in\{1,\,2\}$, with OPEs as in (\ref{Dirac}). Compatibility
with supersymmetry then implies that the superpartners of these fields yield the two $\mathfrak u(1)^4$-current
algebras, as is required in order to identify our theory as a toroidal one.
\hspace*{\fill}\qed
\end{enumerate}
\end{enumerate}

\bibliographystyle{kw}
\bibliography{kw}

\end{document}